\documentclass[aapm, mph, reprint, amsmath, amssymb]{revtex4-1}

\usepackage{graphicx}
\usepackage{tikz}
\usetikzlibrary{calc,shapes}
\usepackage{hyperref}
\usepackage{booktabs} 
\usepackage{color}
\usepackage{amsmath}
\usepackage{cancel}
\usepackage{mathtools}
\usepackage{MnSymbol}
\usepackage{tabularx}
\usepackage{caption}
\usepackage{mathrsfs}
\usepackage{comment}
\usepackage{float}     % Per posizionare le figure esattamente dove vuoi
\usepackage{xcolor}
\usepackage[normalem]{ulem}

\usepackage{dcolumn}% Align table columns on decimal point
\usepackage{bm}% bold math

\usepackage{natbib} % Supporto per BibTeX

\usepackage[mathlines]{lineno}% Enable numbering of text and display math
\modulolinenumbers[5]% Line numbers with a gap of 5 lines
\begin{document}

\preprint{APS/123-QED}

\title{$A=2,3,4$ nuclear contact coefficients in the generalized contact formalism}
%\thanks{Footnote to title of article.}

\author{E. Proietti$^{1,2}$}
\email{eleonora.proietti@phd.unipi.it}
\author{L. E. Marcucci$^{1,2}$}
\email{laura.elisa.marcucci@unipi.it}
\author{M. Viviani$^{2}$}
\email{michele.viviani@pi.infn.it}

\affiliation{$^1$Dipartimento di Fisica "E. Fermi", Università di Pisa, Largo Pontecorvo 3, 56127 Pisa, Italy.}
\affiliation{$^2$Istituto Nazionale di Fisica Nucleare, Sezione di Pisa, Largo Pontecorvo 3, 56127 Pisa, Italy.}

\begin{abstract}
We present a theoretical calculation for the $A = 2, 3$ and 4 nuclear contact coefficients within the generalized contact formalism, using both local and non-local chiral potentials. The Hyperspherical Harmonics method is employed to calculate the nuclear wave functions, from which we derive two-body momentum distributions and density functions to extract the contact coefficients.
    We have extracted the contact coefficients from two-body momentum distributions or from density functions, for a given nucleus and potential, and we have found that the generalized contact formalism predictions are verified in the triplet spin channel for local and non-local potentials. On the other hand, some significant tensions exist for the singlet channels, especially when studied with non-local potentials. We have also analyzed the model-independence of the ratios between the contact coefficients, and we have found to be quite satisfied. This study extends previous works based on local interaction models only.
\end{abstract}

\maketitle

\section{Introduction}
\label{sec:intro}

The study of short-range correlations (SRCs) in nuclei is essential for understanding the nucleon behavior at high momenta or equivalently short distances, where traditional mean-field approaches, such as the nuclear shell model, fail to account for the complex dynamics of nucleon-nucleon correlations.
In this work, we take the operative definition of SRCs as in Refs.~\cite{RevModPhys.89.045002,CiofidegliAtti:2015lcu}, i.e.\ a two-nucleon pair having small center-of-mass momentum and large relative momentum, compared with the typical nuclear Fermi momentum ($ k_F \approx 250 \, \text{MeV}/c $). These pairs are predominantly neutron-proton pairs.

Definitive experimental evidence for SRCs has been obtained through two types of high-energy reactions. The first is the inclusive $ (e, e') $ scattering at Bjorken $ x_B > 1 $, which provides insights into the high-momentum tail of the nuclear momentum distribution~\cite{PhysRevC.48.2451,CLAS:2003eih, PhysRevLett.108.092502}. The second involves exclusive reactions that isolate the effects of ground-state correlations, distinguishing them from competing processes such as two-body 
meson-exchange currents and final-state interactions~\cite{PhysRevLett.90.042301, Piasetzky:2006ai, Hen:2014nza}. Moreover, $ (e, e') $ reactions at large values for $ x_B $ reveal that all nuclei exhibit similar momentum distributions at high momenta, providing direct evidence for the existence of strongly correlated two-nucleon clusters in the nuclear ground state. Further experimental evidence shows that SRC pairs, predominantly consisting of neutron-proton pairs, account for a significant fraction of nucleons in the nucleus at high momenta~\cite{Piasetzky:2006ai,Hen:2014nza, PhysRevLett.113.022501}.

The presence of SRCs in nuclei leads to significant effects on various nuclear phenomena, including the high-momentum tails of nuclear momentum distributions~\cite{RevModPhys.89.045002}, the internal structure of nucleons within nuclei~\cite{RevModPhys.89.045002, PhysRevLett.106.052301, PhysRevC.85.047301, Hen:2013oha, CLAS:2019vsb}, nuclear charge radii~\cite{PhysRevC.75.051303, PhysRevC.76.024315, Menendez:2008jp, PhysRevC.79.055501, PhysRevC.90.065504, Cruz-Torres:2019fum, Wang:2019hjy}, neutrinoless double-beta decay~\cite{PhysRevC.75.051303}, and the properties of neutron stars~\cite{Li:2018lpy, PhysRevC.91.025803, Frankfurt:2008zv}. Additionally, they play a critical role in the understanding of the European Muon Collaboration (EMC) effect, challenging our understanding of how SRCs reconcile with high-momentum transfer reactions.

In this work, we want to test the genaralize contact formalism (GCF), previously used to describe SRCs~\cite{PhysRevC.92.054311, WEISS2018211,Cruz-Torres:2019fum}. In this formalism, the nuclear wave function at high momenta is factorized into a strongly interacting nucleon pair wave function and a term involving the residual system. This factorization, if valuable, would have the advantage of significantly simplifying the calculation of SRCs, by allowing a clear separation between the correlated two-body nucleon pair and the uncorrelated residual system. The idea of writing the nuclear wave function as a product of the two-body term and the residual one is in fact older than the development of the GCF, as, for instance, it is present also in Ref.~\cite{FRANKFURT1981215}. By focusing on the two-body pair, the GCF links two-body density functions (2BDFs) or momentum distributions (2BMDs) to the square of the correlated pair wave function, or its Fourier transform. The link is represented by a coefficient, called contact coefficient, which in this formalism can be seen as a direct measure of the probability of finding nucleon pairs in specific spin-isospin states, providing valuable insight into the short-range structure of the nuclear wave function~\cite{PhysRevC.92.054311, Cruz-Torres:2019fum}. As a result, these coefficients could become an essential tool for quantifying SRCs in different nuclei, offering a connection between theoretical predictions and experimental observables~\cite{WEISS2018211}.

According to the GCF, the contact coefficients for a given nucleus and potential should be the same whether extracted from the high-momentum tail of 2BMDs or the short-distance region of 2BDFs, as the physics governing high momenta and short distances should be universal. This has been verified in Ref.~\cite{Cruz-Torres:2019fum}, albeit using only local models for the nuclear interaction. Furthermore, in Ref.~\cite{Cruz-Torres:2019fum} it has been shown that the ratio of contact coefficients between different nuclei is largely independent of the nuclear interaction employed for a given nucleus. This suggests that SRCs are primarily driven by the fundamental properties of the nuclear force and share similar features across different nuclei, independently of the specific interaction model used.

While previous studies have focused mainly on local interaction potentials, the inclusion of non-local chiral interactions is crucial for a comprehensive understanding of SRCs. In this work, we extend the analysis of SRCs in light nuclei ($ A = 2, 3, 4 $) working in chiral effective field theory (ChEFT) and using both local and non-local chiral interactions. In order to calculate the nuclear wave functions from which 2BMDs and 2BDFs are derived, we have used the Hyperspherical Harmonics (HH) method~\cite{Marcucci:2019hml, Viviani:2005gu}, which is able to solve the $A$-body bound states with both local and non-local chiral potentials. This enables us to investigate the model-dependence of the extracted contact coefficients with a much larger variety of nuclear potentials compared with Ref.~\cite{Cruz-Torres:2019fum}. 

This paper is organized as follows. In Sec.~\ref{sec:theoform} we briefly review the theoretical formalism, and in particular the definition of 2BMDs and 2BDFs, a discussion of the GCF, and the definition of the contact coefficients. In Sec.~\ref{sec:coeff-extract} we explain how contact coefficients are extracted from 2BMDs and 2BDFs. In Sec.~\ref{sec:res} we present our results, and in Sec.~\ref{sec:conc-out} we conclude with a summary and an outlook.

\section{Theoretical Formalism}
\label{sec:theoform}
\subsection{Two-body momentum distributions and density functions}
\label{subsec:2BDFMD}

The probability of finding two nucleons, $N_1$ and $N_2$, with center-of-mass momentum $K$ and relative momentum $k$ can be written as
\begin{equation}
        n_{N_1N_2}({k},{K}) = \int d\hat{\bm{k}} \int d\hat{\bm{K}} \; \psi^\dagger(\bm{k},\bm{K}) \, P_{N_1 N_2} \, \psi(\bm{k},\bm{K})\ , \label{eq:2bmd-def}
\end{equation}
where $\psi(\bm{k},\bm{K})$ is the $A$-nucleus wave function obtained using the HH method~\cite{Viviani:2005gu, Kievsky:2008es, Marcucci:2018llz}, and $P_{N_1 N_2}$ is the projection operator on the nucleon pair $N_1 \, N_2$ = $pn,\, pp,\, nn$. 

 By integrating over the center-of-mass momentum, one obtains the 2BMD, that depends only on the relative momentum $k$, and is given by
\begin{eqnarray}
    n_{N_1N_2}({k}) &=& 4\pi \int dK \; K^2 \, n_{N_1 N_2}({k},{K}) \nonumber \\
    &=&  \! \int d\hat{\bm{k}} \! \int d{\bm{K}} \; \psi^\dagger(\bm{k},\bm{K}) \, P_{N_1 N_2} \, \psi(\bm{k},\bm{K})\ . \label{2bmd_NN}
\end{eqnarray}
If in alternative to $P_{N_1 N_2}$ we use the projection operator
$P^{ST}$, which projects on the pair spin-isospin $ST$ state, we define the spin-isospin 2BMD as
\begin{equation}
    n^{ST}({k}) = \int d \hat{\bm{k}} \int d \bm{K} \; \psi^{\dagger}(\bm{k},\bm{K}) P^{ST}\psi(\bm{k},\bm{K})\ . \label{2bmd_ST}
\end{equation}
Finally, we can join the two projection operators, and projects on the nucleon pair $ N_1 N_2 $ with relative momentum $ k $, in a given spin state $ S $. In this case, we define
\begin{eqnarray}
n_{N_1\, N_2}^S({k}) &=& \int d \hat{\bm{k}} \int d \bm{K} \; \psi^{\dagger}(\bm{k},\bm{K}) P_{N_1\, N_2} P^S \psi(\bm{k},\bm{K}) \nonumber\\
& =& \int d \hat{\bm{k}} \int d \bm{K} \; \psi^{\dagger}(\bm{k},\bm{K})  P_{N_1\, N_2}^S \psi(\bm{k},\bm{K})\label{2bmd_NNS}\ .
\end{eqnarray}

Working in coordinate space, analogous definitions can be formulated for the 2BDFs. In particular, for two nucleons $N_1$ and $N_2$ with relative distance $r$, the 2BDF is given by
\begin{equation}
    \rho_{N_1N_2}({r}) = \int d\hat{\bm{r}} \int d \bm{R}  \, \psi^\dagger(\bm{r},\bm{R}) \, P_{N_1 N_2} \, \psi(\bm{r},\bm{R})\ ,\label{2bdf_NN}
\end{equation}
where we have indicated with ${\bm{R}}$ the $N_1 N_2$ pair center-of-mass position and with ${\bm{r}}$ the relative distance. Similarly, for two nucleons in the spin-isospin state $ST$, the 2BDF can be expressed as 
\begin{equation}
\rho^{ST}({r}) = \int d \hat{\bm{r}} \int d \bm{R} \; \psi^{\dagger}(\bm{r},\bm{R}) P^{ST}\psi(\bm{r},\bm{R})\ ,\label{2bdf_ST}
\end{equation}
while for two nucleons $ N_1 N_2 $ in a given spin state $ S $, the 2BDF can be expressed as
\begin{equation}
\rho_{N_1\, N_2}^S({r})  = \int d \hat{\bm{r}} \int d \bm{R} \; \psi^{\dagger}(\bm{r},\bm{R}) P_{N_1\, N_2}^S\psi(\bm{r},\bm{R})\ . \label{2bdf_NNS}
\end{equation}
A crucial point now is the calculation of the nuclear wave functions
$\psi(\bm{r},\bm{R})$ and $\psi(\bm{k},\bm{K})$, i.e.\ the adopted model of the nuclear interaction and the {\it ab-initio} method used to solve the $A$-body quantum problem. For the latter, we have used the HH method, which has the advantage of being a highly accurate method working both in coordinate- and in momentum-space. This versatility is crucial for allowing us to use both local and non-local chiral interactions. 

The HH method has been extensively discussed for the $A=3,4$ bound states in Refs.~\cite{Kievsky:2008es,Marcucci:2019hml,Viviani:2005gu}. Here we briefly summarize the main steps of the HH technique, necessary to understand the details of the present work. The $A$-body wave function is expanded in terms of some known basis functions (the spin-isospin HH functions) as
\begin{equation}
\psi=\sum_\mu c_\mu \phi_\mu \ , \label{eq:HHexp}
\end{equation}
where
\begin{equation}
    \phi_\mu=u_\mu(\rho)\,Y_\mu(\Omega_\rho)\ ,
    \label{eq:phimu-r}
\end{equation}
or
\begin{equation}
    \phi_\mu=w_\mu(Q)\,Y_\mu(\Omega_Q)\ ,
    \label{eq:phimu-q}
\end{equation}
depending whether we are working in coordinate- or in momentum-space. In the above equations we have indicated with $Y_\mu(\Omega_\rho)$ or 
$Y_\mu(\Omega_Q)$ the spin-isospin HH functions in $r$- or $k$-space, and are constructed in such a way that the wave function is fully anti-symmetric under the exchange of any pair of nucleons. Furthermore, $u_\mu(\rho)$ or $w_\mu(Q)$ are the hyperradial functions or their Fourier transform, and therefore function of the hypermomentum $Q$. In $r$-space, $u_\mu(\rho)$ is chosen to be 
\begin{equation}
u_\mu(\rho)\propto \mathcal{L}^{(3A-4)}_k(\gamma \rho) e^{-\gamma \rho / 2}\ ,
\label{eq:laguerre}
\end{equation}
where $\mathcal{L}^{(3A-4)}_k(\gamma \rho)$ is a Laguerre polynomial and $\gamma$ a non-linear parameter, typically chosen between 2.5–4.5 fm$^{-1}$ for local potentials and between 4–8 fm$^{-1}$ for non-local potentials. 
The not-explicitly given proportionality factor is chosen to make $u_\mu(\rho)$ orthonormal.
The expansion coefficients $ c_\mu$ of Eq.~\eqref{eq:HHexp} are determined by the Rayleigh-Ritz variational principle, which leads to the solution of a set of coupled differential equations given by
\begin{equation}
\sum_{\mu'} c_{\mu'} \langle \phi_\mu | H - E | \phi_{\mu'} \rangle = 0\ ,
\label{eq:RRvar}
\end{equation}
where $E$ is the energy of the state. This system of equations can be solved with standard numerical techniques.

As already remarked, the HH method can be used with essentially any type of two-nucleon (NN) potential, both local and non-local. However, at present, the method can be used with only local three-nucleon interactions (TNI). Even so, this leaves a large flexibility on the choice of the nuclear interaction, and we list here those considered in the present work.

The first interaction model adopted is the phenomenological Argonne $v_{18}$ (AV18) NN interaction~\cite{Wiringa:1994wb}, complemented by the Urbana IX (UIX) TNI~\cite{PhysRevLett.74.4396}. The use of the AV18/UIX model has allowed us to validate our results against the ones in Ref.~\cite{Cruz-Torres:2019fum}. Additionally, we have employed various chiral potentials.
In particular, we have used the local Norfolk chiral potentials (NV), both without and with associated TNI~\cite{PhysRevC.91.024003, PhysRevC.98.044003}. We have focused on both those of class I and II, depending on the range of laboratory energy employed in the fitting procedure ($E_{LAB}=125$ MeV or 200 MeV, respectively), with various combinations for the short- ($R_S$) and long-range ($R_L$) cutoffs (model a with $(R_S,R_L)=(0.7,1.0)$ fm, and model b with $(R_S,R_L)=(9.8,1.2)$ fm, respectively). The adopted NV potentials have been labelled NV2+3/Ia*, NV2+3/Ib*, NV2+3/IIa*, and NV2+3/IIb*. Note that the star indicates that the parameters entering in the TNI, i.e.\ the low-energy constants (LECs) $c_D$ and $c_E$, have been fitted to the triton binding energy and half-life~\cite{PhysRevC.98.044003}. 

Furthermore, we have considered the non-local chiral NN interactions from Ref.~\cite{PhysRevC.96.024004}, derived at next-to-next-to leading order (N2LO) and higher, up to N4LO. Different high-momentum cutoffs have been considered, i.e.\ $\Lambda=450,500,550$ MeV. We have started from N2LO, as this is the first order at which TNI appears. Also in conjunction with these non-local NN interactions, we have used a local TNI, with the LECs again fitted to the triton binding energy and half-life~\cite{Gnech:2023mvb}. These non-local interactions have been labelled according to the chiral order and the cutoff value. For instance, N2LO500 indicates the potential at N2LO, with cutoff $\Lambda=500$ MeV.
We remark that the reason behind the locality of the adopted TNI models resides on the fact that at present the HH method has not been implemented for non-local TNI models. 

All the adopted models are able to nicely describe the $ A = 3, 4$ nuclei properties, as binding energies, charge or matter radii, and also $A=3,4$ scattering observables. 

\subsection{Generalized contact formalism}
\label{subsec:GCF}

We review the GCF in this subsection. The details of this formalism can be found, for instance, in Refs.~\cite{PhysRevC.92.054311,WEISS2018211}.

In the GCF, the nuclear wave function $ \psi $ is factorized at short distances into a strongly interacting nucleon pair and a weakly interacting residual system. Therefore, we can write~\cite{PhysRevC.92.054311}
\begin{equation}
    \psi \xrightarrow{r_{ij} \rightarrow 0} \sum_\alpha \varphi_{ij}^\alpha(\bm{r}_{ij}) \, A^\alpha_{ij}(\bm{R}_{ij},{\bm{r}_{k \neq ij}})\ , \label{expansion}
\end{equation}
where $ij = nn, np, pp$, and $\alpha$ refers to the pair quantum numbers $L,S,J$, $L$ being the orbital angular momentum, $S=0,1$ the pair spin, and ${\bm J}={\bm L}+{\bm S}$ being the total angular momentum. The limit $r_{ij}\to 0$ can be quantified as $r_{ij} \lesssim 1$ fm. The function $A_{ij}^\alpha(\bm{R}_{ij},{\bm{r}_{k \neq ij}})$ describes the residual $ A - 2 $ system, while the term $ \varphi_{ij}^\alpha(\bm{r}_{ij}) $ represents the two-body wave function, and it is referred to as the universal function, because is expected not to depend on the nucleus, but only on the distance between the two nucleons, the set of quantum numbers $ \alpha $, and the NN interaction.

Generally, the expansion in Eq.~\eqref{expansion} may involve more than one term per channel. However, in this work, only the dominant SRC-contributing channels are considered. These include the $ pn $ deuteron channel ($ L = 0, 2 $, $ S = 1 $, $ J = 1 $) and the singlet $ nn $, $ np $, $ pp $ $ s $-wave channels ($ L = S = J = 0 $). This simplification is justified since, in the limit $ r_{ij} \to 0 $, contributions from other channels are  expected to be negligible.

The two-body universal function $\varphi^\alpha_{ij}({\bm r}_{ij})$ is written as
\begin{equation}
    \varphi^\alpha_{ij}({\bm r}_{ij}) = [\varphi_{ij}^{\{S \, J\} \, L}(\bm{r}_{ij}) \otimes \chi_{S \, T}]_{J \, M}\ ,
\label{eq:phi-uf}
\end{equation}
where $ \varphi_{ij}^{\{S \, J \} \, L}(\bm{r}_{ij}) $ is the spatial part of the wave function, $ \chi_{S \, T} $ is the spin-isospin part, and $ J, \, M $ are the total angular momentum of the pair and its projection, respectively. The spatial component $\varphi_{ij}^{\{S \, J\} \, L}(\bm{r}_{ij})$ is expressed as a product of radial and angular parts as
\begin{equation}
    \varphi_{ij}^{\{S \, J\} \, L}(\bm{r}_{ij}) = \varphi^{L \, S \, J}_{ij}(r_{ij}) \, Y_{ L \, L_z}(\hat{\bm r}_{ij})\ ,
    \label{eq:phiLSJ}
\end{equation}
where $ \varphi^{L \, S \, J}_{ij}(r_{ij}) $ is the radial wave function and $ Y_{L \, L_z}(\hat{\bm r}_{ij}) $ is the spherical harmonic with orbital angular momentum of the pair $ L $ and projection $L_z$. The radial function depends solely on the relative pair distance $ r_{ij} $.
For $ S = 0 $ channels, $ \varphi^{L \, S \, J}_{ij}(r_{ij}) $ is obtained as the zero-energy two-body scattering wave function with  $ L = S = J = 0 $. For the $ S = 1 $ channel, it is chosen to be the deuteron ground state ($ L = 0, 2 $, $ S = 1 $, $ J = 1 $).

When working in momentum space, it has been shown in Ref.~\cite{PhysRevC.92.054311} that, within the GCF, the nuclear wave function in momentum space, for large relative momentum $k_{ij}$, can be approximated as
\begin{equation}
    \psi \xrightarrow{k_{ij} \to \infty} \sum_\alpha \tilde{\varphi}_{ij}^\alpha(\bm{k}_{ij}) \, A^\alpha_{ij}(\bm{K}_{ij},{\bm{k}_{k \neq ij}})\ , \label{expansion-p}
\end{equation}
where the universal function $\tilde{\varphi}_{ij}^\alpha(\bm{k}_{ij})$ is described as in Eqs.~\eqref{eq:phi-uf} and~\eqref{eq:phiLSJ}, but with $\bm{r}_{ij}$ replaced with  $\bm{k}_{ij}$. In order to calculate $\tilde{\varphi}_{ij}^\alpha(\bm{k}_{ij})$, the Fourier transform (FT) 
of the coordinate-space solution $ \varphi^{L \, S \, J}_{ij}(r_{ij}) $ is required. This presents no challenge for the $S=1$ channels. Instead, 
for the $ S = 0 $ channels, the FT of the zero-energy scattering state wave function would not converge. To resolve this, a Gaussian cutoff $ e^{-(r_{ij}/c)^2} $, with $ c \sim 50 \, \mathrm{fm} $, is applied to suppress large-distance contributions~\cite{PhysRevC.92.054311,WEISS2018211}. This regularization preserves the high-momentum behavior, corresponding to small distances, ensuring that the result remains independent of $ c $ as $ k \to \infty $.

The choice of the spin $ S $ for a given nucleon pair $ N_1 N_2 $ also fixes the other quantum numbers ($ L $ and $ J $). Therefore, in the following, instead of considering the dependence on $ \alpha $, we will explicitly focus on the dependence only on $ S $. Hereafter, we will replace $ ij $ with $ N_1 \, N_2 $ to denote the nucleon pair, and we will use $ r $ and $k$, with no subscripts, to represent the relative distance and momentum of the nucleon pair.

\subsection{Nuclear contact coefficients}
\label{subsec:NCC}

Since SRCs appear in the 2BMDs and 2BDFs at large values of $k$, or equivalently small values of $r$, we focus on these $k$ and $r$ regimes. Therefore, 2BMDs and 2BDFs should exhibit scaling with 
$ \tilde{\varphi}^S_{N_1 \, N_2}(k) $ and $ {\varphi}^S_{N_1 \, N_2}(r) $, respectively, when they are calculated using the same potential in a given channel (i.e.\ for the same $ N_1, N_2 $, and $ S $).
As a consequence, we can write
\begin{eqnarray}
   n_{N_1 N_2, \; A}^S({k}) & \xrightarrow{k \to \infty}& \tilde{C}_{N_1 N_2, \; A}^S \, \times \,  |\tilde{\varphi}^S_{N_1 N_2}(k)|^2\ , \label{2bmd} \\
    \rho_{N_1 N_2, \; A}^S(r) & \xrightarrow{r \to 0} & {C}_{N_1 N_2, \; A}^S \, \times \,  |\varphi^S_{N_1 N_2}(r)|^2\ , \label{2bdf} 
\end{eqnarray}
where the coefficients ${C}_{N_1 N_2, \; A}^S $ and $\tilde{C}_{N_1 N_2, \; A}^S$ are the nuclear contact coefficients. Note that while the functions $\varphi^S_{N_1 N_2}(r)$ and $\tilde{\varphi}^S_{N_1 N_2}(k)$ are universal, i.e.\ depend solely on the nuclear interaction in the given spin channel $S$, but not on the considered nucleus, the coefficients ${C}_{N_1 N_2, \; A}^S $ and $\tilde{C}_{N_1 N_2, \; A}^S$ depend on the nuclear interaction, the spin channel $S$, and the considered nucleus. This is the reason for which we have added the subscript $A$ to both the 2BMDs or 2BDFs, and to the nuclear coefficients.
Eqs.~\eqref{2bmd} and~\eqref{2bdf} are used in order to quantitatively address the $k\to\infty$ and $r\to 0$ limits. Operatively, we will consider the ranges of $k$ and $r$ where a plateau in the ratio $n_{N_1 N_2, \; A}^S({k})/|\tilde{\varphi}^S_{N_1 N_2}(k)|^2$ and $\rho_{N_1 N_2, \; A}^S(r)/|\varphi^S_{N_1 N_2}(r)|^2$ is found. This is similar to what has been done in Ref.~\cite{Cruz-Torres:2019fum}. Note that in our approach we do not consider three-nucleon SRCs. These could be present in $^4$He. Although a dedicated study on this is of clear interest, it is beyond the scope of this work. Furthermore, we anticipate that the plateau in the ratio $n_{N_1 N_2, \; A}^S({k})/|\tilde{\varphi}^S_{N_1 N_2}(k)|^2$ is always below $k\sim 5$ fm$^{-1}$. Therefore, the two-nucleon relative energy is of the order of half of the nucleon mass. Although in such an energy regime non-relativistic approaches could be considered questionable, they are known to work quite well, especially in electron scattering. As an example, we mention the nice description of light nuclei electric and magnetic form factors at momentum transfer up to 5 fm$^{-1}$ (see, e.g., the review of Ref.~\cite{Marcucci_2016}). 

We finally remark that in the GCF, the construction that links short-distance and high-momentum physics leads to the expectation that the contact coefficients satisfy the relation $ \tilde{C}_{N_1 N_2, \, A}^S = C_{N_1 N_2, \, A}^S $\cite{WEISS2018211}. In this work, we will verify whether this relation, predicted by the GCF, holds for all spin-isospin channels, for all potentials, both local and non-local, and for each considered nucleus.

\section{Extraction of the nuclear contact coefficients}\label{sec:coeff-extract}

In this work, we focus on the contact coefficients $ C_{nn/pp, \; A}^{S=0} $, $ C_{np, \; A}^{S=1} $, $ C_{np, \; A}^{S=0} $, and $ \tilde{C}_{nn/pp, \; A}^{S=0} $, $ \tilde{C}_{np, \; A}^{S=1} $, $ \tilde{C}_{np, \; A}^{S=0} $ for $ A = 2, 3, 4 $. By assuming the Coulomb interaction to be negligible respect to the nuclear interaction, we can impose $ C_{nn, \; A}^{S=0} \equiv C_{pp, \; A}^{S=0} $ and $ \tilde{C}_{nn, \; A}^{S=0} \equiv \tilde{C}_{pp, \; A}^{S=0} $.
According to Eqs.~\eqref{2bmd} and~\eqref{2bdf}, we extract the contact coefficients from the ratios $ n(k) / |\tilde{\varphi}(k)|^2 $ for large values of $ k $, and $ \rho(r) / |\varphi(r)|^2 $ for small values of $ r $, where we have dropped here for simplicity the subscripts and superscripts. From the same equations, we expect in the large-$k$ and small-$r$ regions to see a plateau for these ratios. 

As a first step, we have adopted a procedure similar to the one of Ref.~\cite{Cruz-Torres:2019fum}. We should first of all notice that in the definitions of the contact coefficients given in Eqs.~\eqref{2bmd} and~\eqref{2bdf}, only the spin channel $S$ is fixed. However, in Ref.~\cite{Cruz-Torres:2019fum} the 2BMDs and 2BDFs of Ref.~\cite{PhysRevC.89.024305} were used, obtained within the Variational Monte Carlo (VMC) approach. The available 2BMDs or 2BDFs of Ref.~\cite{PhysRevC.89.024305} are $n_{N_1 N_2}(k)$ or $\rho_{N_1 N_2}(r)$ and  $n^{ST}(k)$ or $\rho^{ST}(r)$. No 2BMDs or 2BDFs are available with a fixed $S$ value (not $ST$). Then, in order to extract the contact coefficients, the assumption of $L=0$, $L$ being the pair orbital angular momentum, is applied. Therefore, the $S=0$ ($S=1$) state corresponds to the $T=1$ ($T=0$) state. With this approximation, the contact coefficients are defined as~\cite{Cruz-Torres:2019fum}
\begin{eqnarray}
    \tilde{C}_{pp/nn}^{S=0}(k) & = & \lim\limits_{k \to \infty} \, \left\{\frac{n_{pp/nn}({k})} {|\tilde{\varphi}_{pp/nn}^{S=0}({k})|^2}\right\}\ , \label{eq:cpp0_old_k} \\
    \tilde{C}^{S=1}_{pn}(k) & = & \lim\limits_{k \to \infty} \, \left\{ \frac{n^{ST=10}({k})}{|\tilde{\varphi}_{pn}^{S=1}({k})|^2}\right\}\ , \label{eq:cpn1_old_k}\\
    \tilde{C}^{S=0}_{pn}(k) & = &\lim\limits_{k \to \infty} \, \left\{ \frac{n_{pn}({k})  - \tilde{C}^{S=1}_{pn} |\tilde{\varphi}_{pn}^{S=1}({k})|^2}{|\tilde{\varphi}_{pn}^{S=0}({k})|^2}\right\}\ .
    \label{eq:cnp0_old_k}
\end{eqnarray}
Similarly, in coordinate space we have 
\begin{eqnarray}
     C_{pp/nn}^{S=0}(r) & = & \lim\limits_{r \to 0} \, \left\{\frac{\rho_{pp/nn}({r})}{|\varphi_{pp/nn}^{S=0}({r})|^2}\right\}\ , \label{eq:cpp0_old_r} \\
     C^{S=1}_{pn}(r) & =& \lim\limits_{r \to 0} \, \left\{
     \frac{\rho^{S=1}_{pn}({r})}{|\varphi_{pn}^{S=1}({r})|^2}\right\}\ , \label{eq:cnp1_old_r} \\
     C^{S=0}_{pn}(r) & =& \lim\limits_{r \to 0} \, \left\{ 
     \frac{\rho_{pn}({r}) - C^{S=1}_{pn} |\varphi_{pn}^{S=1}({r})|^2} 
     {|\varphi_{pn}^{S=0}({r})|^2} \right\}\ . \label{eq:cnp0_old_r}
\end{eqnarray}
It should be noticed that in the $pp/nn$ case, being a state with $T=1$, $S=0$ is implied again by the assumption $L=0$. We started our study defining the contact coefficients as in Ref.~\cite{Cruz-Torres:2019fum}, i.e.\ using Eqs.~\eqref{eq:cpp0_old_k}--\eqref{eq:cnp0_old_r}. However, thanks to the definitions of Eqs.~\eqref{2bmd_NNS} and~\eqref{2bdf_NNS}, we have been able also to remove the approximation $L=0$, and therefore to redefine the contact coefficients as
\begin{eqnarray}
    \tilde{C}_{pp/nn}^{S=0}(k)  &=& \lim\limits_{k \to \infty} \, \left\{ \frac{n_{pp/nn}^{S=0}({k})}{|\tilde{\varphi}_{pp/nn}^{S=0}({k})|^2} \right\}\ , \label{eq:cpp0_k} \\
    \tilde{C}^{S=1}_{pn}(k)  &=& \lim\limits_{k \to \infty} \, \left\{ \frac{n^{S=1}_{pn}({k})}{|\tilde{\varphi}_{pn}^{S=1}({k})|^2} \right\}
    \ , \label{eq:cnp0_k} \\
    \tilde{C}^{S=0}_{pn}(k)  &=& \lim\limits_{k \to \infty} \, \left\{ \frac{n^{S=0}_{pn}({k})} {|\tilde{\varphi}_{pn}^{S=0}({k})|^2} \right\}\ , \label{eq:cnp1_k}
\end{eqnarray}
using the 2BMDs, and
\begin{eqnarray}
    C_{pp/nn}^{S=0}(r) & = &\lim\limits_{r \to 0} \, \left\{
    \frac{\rho_{pp/nn}^{S=0}({r})}{|\varphi_{pp/nn}^{S=0}({r})|^2} \right\}\ , \label{eq:cpp0_r} \\
    C^{S=1}_{pn}(r) & = & \lim\limits_{r \to 0} \, \left\{ 
    \frac{\rho^{S=1}_{pn}({r})}{|\varphi_{pn}^{S=1}({r})|^2} \right\}\ , \label{eq:cnp1_r} \\
    C^{S=0}_{pn}(r) & = & \lim\limits_{r \to 0} \, \left\{ 
    \frac{\rho^{S=0}_{pn}({r})} {|\varphi_{pn}^{S=0}({r})|^2} \right\}\ , \label{eq:cnp0_r} 
\end{eqnarray}
using the 2BDFs.
The practical derivation of the contact coefficients is achieved searching for a plateau in the large-$k$ or small-$r$ regions for the ratios between the 2BMDs or 2BDFs and the universal functions $\tilde{\varphi}_{N_1 N_2}^{S}({k})$ or $\varphi_{N_1 N_2}^{S}({r})$, according to Eqs.~\eqref{eq:cpp0_old_k}--\eqref{eq:cnp0_r}. 
Specifically, for the $ k $-space coefficients,  the plateau region for local potentials occurs beyond 3--3.5 fm$^{-1}$, while for non-local potentials, the plateau region is found at slightly smaller values of $ k $, roughly between 2 and 3 fm$^{-1}$. Indeed, due to divergences at high momenta caused by cutoff effects, the corresponding range of $ k $ cannot be considered. To be noticed that our findings for local potentials are the same as those found in previous works on the GCF~\cite{PhysRevC.92.054311,WEISS2018211,Cruz-Torres:2019fum}. In $ r $-space, the plateau region for both local and non-local potentials consistently appears as $ r \to 0 $.
More in detail, the contact coefficients were extracted from the plateau region by averaging values over an interval of $ 1 \, \text{fm} $ in $r-$space or $ 1 \, \text{fm}^{-1} $ in $k-$space.
This interval in $ r $-space was selected in the same way for both local and non-local potentials, centered at $ r = 0.5 \, \text{fm} $. In $ k $-space, for local potentials, the interval was centered at $ k = 4 \, \text{fm}^{-1} $. For non-local potentials, due to the divergences caused by cutoff effects, the plateau region varied depending on the potential, particularly with respect to the order and cutoff. Consequently, the interval was centered at the minimum value within the plateau region to ensure a consistent criterion also for non-local potentials.
We then calculated the uncertainties associated with the extraction of the contact coefficients through the standard deviation, computed over the same interval used for extracting the contact coefficients. The uncertainties were calculated to quantify the reliability of the extracted contact coefficients and account for fluctuations within the selected interval.

As a final step, we have calculated ratios of the extracted contact coefficients of the various nuclei with respect to a reference nucleus. These ratios, by definition, are given by 
the ratios of 2BMDs at large values of $ k $ and of 2BDFs at small values of $ r $, i.e.\
\begin{eqnarray}
   \frac{\rho_{N_1 N_2, \; A}^S(r)}{\rho_{N_1 N_2, \; A_0}^S(r)}
   &\xrightarrow[]{r \rightarrow 0}&
   \frac{{C}_{N_1 N_2, \; A}^S}{{C}_{N_1 N_2, \; A_0}^S}\ , \label{ratio_r}\\ 
   \frac{n_{N_1 N_2, \; A}^S(k)}{n_{N_1 N_2, \; A_0}^S(k)}
   &\xrightarrow[]{k \rightarrow \infty }& \frac{\tilde{C}_{N_1 N_2, \; A}^S}{\tilde{C}_{N_1 N_2, \; A_0}^S}\ , \label{ratio_k} 
\end{eqnarray}
where the reference nucleus, denoted with $ A_0 $, has been chosen to be the deuteron in the case of $ S = 1 $, and $ ^4\text{He} $ in the case of $ S = 0 $.
We would like to remark that
according to the GCF~\cite{WEISS2018211}, the
$ {C}_{N_1 N_2, \; A}^S $ and  $\tilde{C}_{N_1 N_2, \; A}^S $ contact coefficients  should coincide for a given nucleus and potential, indicating that the contact coefficients are independent of whether the analysis is performed in $ k $- or $ r $-space. As a consequence, also the ratio between contact coefficients should be the same if we work in $ k $- or $ r $-space. Moreover, the ratios of contact coefficients relative to a reference nucleus $ A_0 $, as defined in Eqs.~\eqref{ratio_r}--\eqref{ratio_k}, are expected, according to the GCF, to be model-independent, meaning they should not vary with the interaction employed for a given nucleus~\cite{Cruz-Torres:2019fum,WEISS2018211}. 
In Ref.~\cite{Cruz-Torres:2019fum}, this was shown 
within uncertainties across different local potentials. 
The aim of this paper is to verify whether this behavior holds true for a large variety of chiral potential, both local and non-local. 
The importance of this model-independence resides on the fact that the contact coefficients ratios for heavy nuclei can be calculated using soft potentials, making in this way the computational effort lighter. Then, these ratios can be multiplied for 
$ \tilde{C}_{N_1 N_2, \, A_0}^S$ (or $C_{N_1 N_2, \, A_0}^S $), obtained using more complicated harder potentials, and $ \tilde{C}_{N_1 N_2, \, A}^S$ (or $C_{N_1 N_2, \, A}^S $) of the heavy $A$-nucleus is obtained using these potentials, without the need of a direct, probably impossible, calculation. 

\section{Results}
\label{sec:res}

We present in this section our results. First of all, in Sec.~\ref{subsec:2b-uf} we discuss the results obtained for the ratios between the 2BDFs or 2BMDs and the universal functions, as requested in Eqs.~\eqref{eq:cpp0_old_k}--\eqref{eq:cnp0_r}, for both $A=3$ and $A=4$. We then discuss in Secs.~\ref{subsec:A34CC} and~\ref{subsec:A34CC-ratio} the results for the contact coefficients and for the ratios of Eqs.~\eqref{ratio_r} and~\eqref{ratio_k}. Preliminarily, in Sec.~\ref{subsec:num-test}, we discuss the numerical tests performed in order to access the reliability of our predictions.

\subsection{Numerical tests}
\label{subsec:num-test}

The first test we have performed is about the numerical integration present in the definition of the 2BMDs and 2BDFs of Eqs.\eqref{eq:2bmd-def}--~\eqref{2bdf_NNS}.
This integrations has been carried out using the Van der Corput sequence~\cite{vanderCorput1935}, and we have verified the stability of the results as the number of Van der Corput integration points increases. Typically, 50,000 points have been found sufficient to achieve converged results for any nucleus and any potential.

As second numerical check, we have verified that we have reached convergence on the terms of the HH expansion basis. The most critical convergence is that related to the maximum number $N_{Lag}$ of Laguerre polynomials  (see Eq.~\eqref{eq:laguerre}). Specifically, we have calculated the contact coefficients as functions of $N_{Lag}$, as listed in Table~\ref{tab:coefficients3} for $A=3$ and in Table~\ref{tab:coefficients4} for $A=4$. These results are obtained using the N3LO500 potential as an illustrative example; similar behavior was observed for all other potentials. In Table~\ref{tab:coefficients3} the TNI is not included, while in Table~\ref{tab:coefficients4} TNI is also present. The reported results demonstrate that convergence on $N_{lag}$ is achieved, confirming the stability of the extracted coefficients, independently on the presence or absence of TNI.
\begin{table*}[htbp]
\begin{tabular}{|c|c|c|c|c|c|c|c|}
\hline
 &  & \multicolumn{2}{|c|}{$C_{np}^{S=1}$} & \multicolumn{2}{|c|}{$C_{nn/pp}^{S=0}$} & \multicolumn{2}{|c|}{$C_{np}^{S=0}$} \\
\hline
 & $N_{Lag}$ & $r$ & $k$ & $r$ & $k$ & $r$ & $k$ \\
\hline
 & 16 & 4.202 $\pm$ 0.026 & 4.107 $\pm$ 0.207 & 0.441 $\pm$ 0.027 & 0.392 $\pm$ 0.012 & 0.226 $\pm$ 0.013 & 0.202 $\pm$ 0.005 \\
$^3$H & 20 & 4.198 $\pm$ 0.026 & 4.089 $\pm$ 0.150 & 0.441 $\pm$ 0.027 & 0.390 $\pm$ 0.009 & 0.225 $\pm$ 0.013 & 0.201 $\pm$ 0.003 \\
 & 24 & 4.198 $\pm$ 0.026 & 4.085 $\pm$ 0.154 & 0.441 $\pm$ 0.027 & 0.390 $\pm$ 0.009 & 0.226 $\pm$ 0.013 & 0.201 $\pm$ 0.003 \\
 & 28 & 4.198 $\pm$ 0.026 & 4.085 $\pm$ 0.157 & 0.441 $\pm$ 0.027 & 0.390 $\pm$ 0.009 & 0.226 $\pm$ 0.013 & 0.201 $\pm$ 0.003 \\
\hline
 & 16 & 4.118 $\pm$ 0.022 & 4.033 $\pm$ 0.199 & 0.412 $\pm$ 0.021 & 0.375 $\pm$ 0.012 & 0.218 $\pm$ 0.012 & 0.196 $\pm$ 0.006 \\
$^3$He & 20 & 4.113 $\pm$ 0.022 & 4.011 $\pm$ 0.142 & 0.411 $\pm$ 0.021 & 0.374 $\pm$ 0.007 & 0.217 $\pm$ 0.012 & 0.195 $\pm$ 0.003 \\
 & 24 & 4.112 $\pm$ 0.022 & 4.009 $\pm$ 0.147 & 0.411 $\pm$ 0.021 & 0.373 $\pm$ 0.007 & 0.217 $\pm$ 0.012 & 0.195 $\pm$ 0.003 \\
 & 28 & 4.112 $\pm$ 0.022 & 4.008 $\pm$ 0.150 & 0.411 $\pm$ 0.021 & 0.373 $\pm$ 0.007 & 0.217 $\pm$ 0.012 & 0.195 $\pm$ 0.003 \\
\hline
\end{tabular}
\caption{Contact coefficients $C_{np}^{S=1}$, $C_{nn/pp}^{S=0}$, $C_{np}^{S=0}$ as functions of the maximum number of Laguerre polynomials $N_{Lag}$ for $^3$H and $^3$He. The N3LO500 potential without TNI is used. Both 2BMDs and 2BDFs are used for the extraction. We use the notation $k$ and $r$ for the two cases, respectively, rather than $\tilde{C}^S_{N_1 N_2}$ and $C^S_{N_1 N_2}$.}
\label{tab:coefficients3}
\end{table*}
\begin{table*}[htbp]
\begin{tabular}{|c|c|c|c|c|c|c|c|}
\hline
 &  & \multicolumn{2}{|c|}{$C_{np}^{S=1}$} & \multicolumn{2}{|c|}{$C_{nn}^{S=0}$} & \multicolumn{2}{|c|}{$C_{np}^{S=0}$} \\
\hline
  & $N_{Lag}$ & $r$ & $k$ & $r$ & $k$ & $r$ & $k$ \\
\hline
 & 16 & 7.461 $\pm$ 0.050 & 8.734 $\pm$ 1.040 & 0.418 $\pm$ 0.020 & 0.401 $\pm$ 0.014 & 0.430 $\pm$ 0.017 & 0.442 0.020
 \\
$^4$He & 22 & 7.462 $\pm$ 0.050 & 8.713 $\pm$ 1.05 & 0.418 $\pm$ 0.020 & 0.396 $\pm$ 0.016 & 0.430 $\pm$ 0.017 & 0.438 $\pm$ 0.016 \\
 & 24 & 7.462 $\pm$ 0.050 & 8.713 $\pm$ 1.06 & 0.418 $\pm$ 0.020 & 0.395 $\pm$ 0.017 & 0.430 $\pm$ 0.017 & 0.437 $\pm$ 0.015 \\
\hline
\end{tabular}
\caption{Same as Table~\ref{tab:coefficients3} but for $^4$He and including TNI.}
\label{tab:coefficients4}
\end{table*}
From inspection of Tables~\ref{tab:coefficients3} and~\ref{tab:coefficients4}, it is also evident that the contact coefficient $ C_{np}^{S=1} $ is approximately an order of magnitude larger than the two $S=0$ coefficients across all analyzed nuclei. Even if the tables report results obtained with the N3LO500 potential, a similar behavior has been observed for all the considered potentials, both local and non-local. This confirms that the $ np$ $S=1 $ channel is the dominant one. 

The second most critical convergence on terms of the HH expansion is that related to the so-called grand-angular momentum ( here denoted with $G$), proportional to the order of the Jacobi polynomial present in the HH function (see Ref.~\cite{Marcucci:2019hml}). 
The final results were calculated with a maximum value for $G$ equal to $ G_{max} = 80 $. In order to verify the independence of the contact coefficients from this choice, additional calculations were performed with $ G_{max} = 100 $. As an example, we report in Table~\ref{tab:G_variation} the contact coefficients for $ ^3\mathrm{He} $, calculated with the N3LO500 and N4LO500 potentials, for $ G_{max} = 80 $ and $ G_{max} = 100 $.
These coefficients are presented with 5 decimal digits to emphasize the stability of the results as $ G $ increases.
\begin{table*}[htbp]
\begin{tabular}{|c|c|c|c|c|c|c|c|}
\hline
 &  & \multicolumn{2}{|c|}{$C_{np}^{S=1}$} & \multicolumn{2}{|c|}{$C_{pp}^{S=0}$} & \multicolumn{2}{|c|}{$C_{np}^{S=0}$} \\
\hline
 $G_{max}$ & Potential & $r$ & $k$ & $r$ & $k$ & $r$ & $k$ \\
\hline
 80 & N3LO500 & 4.08440 $\pm$ 0.00394 & 4.45673 $\pm$ 0.28931 & 0.38042 $\pm$ 0.00784 & 0.41253 $\pm$ 0.01528 & 0.20305 $\pm$ 0.00453 & 0.21780 $\pm$ 0.00873 \\
   & N4LO500 & 4.19681 $\pm$ 0.05137 & 4.25194 $\pm$ 0.14500 & 0.31339 $\pm$ 0.00159 & 0.36688 $\pm$ 0.01598 & 0.16817 $\pm$ 0.00144 & 0.20012 $\pm$ 0.01115 \\
\hline
 100  & N3LO500 & 4.08439 $\pm$ 0.00394 & 4.45671 $\pm$ 0.28929 & 0.38042 $\pm$ 0.00784 & 0.41252 $\pm$ 0.01528 & 0.20305 $\pm$ 0.00453 & 0.21780 $\pm$ 0.00873 \\
    & N4LO500 & 4.19681 $\pm$ 0.05137 & 4.25192 $\pm$ 0.14502 & 0.31339 $\pm$ 0.00158 & 0.36687 $\pm$ 0.01598 & 0.16817 $\pm$ 0.00144 & 0.20012 $\pm$ 0.01114 \\ 
\hline
\end{tabular}
\caption{Contact coefficients $C_{np}^{S=1}$, $C_{pp}^{S=0}$, and
$C_{np}^{S=0}$ for $ ^3\mathrm{He} $ as a function of the maximum value for the grand-angular momentum $ G_{max} $. The N3LO500 and N4LO500 potentials are used.}
\label{tab:G_variation}
\end{table*}
Similar checks were performed for all potentials and for all considered nuclei. In all cases, convergence was reached with $G_{max}=80$ for $A=3$ and $G_{max}=48$ for $A=4$.

\begin{figure}[htbp]
\includegraphics[width=1.0\linewidth]{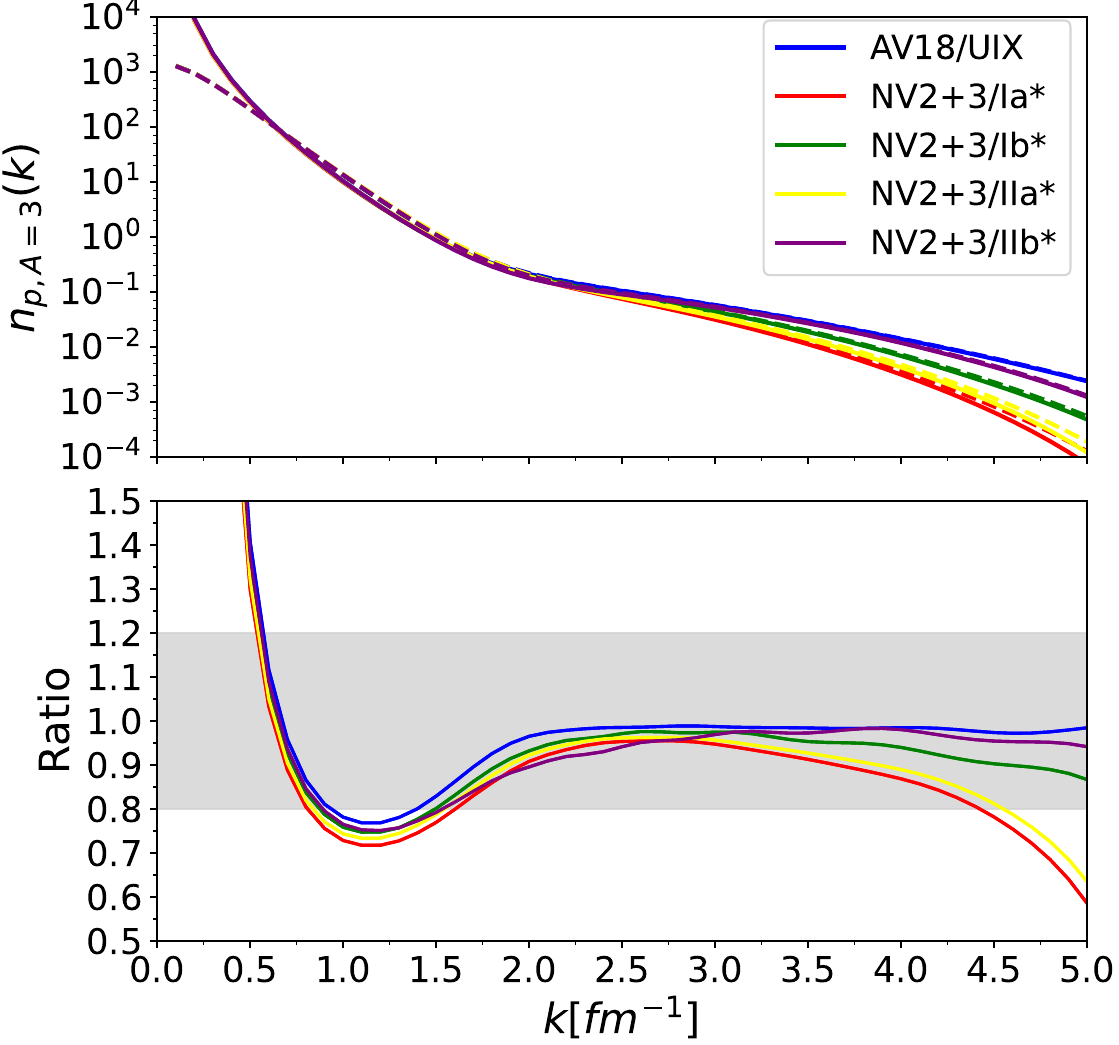}
\caption{1BMDs $ n_{p, A=3}(k) $ (dashed line) and $ n_{p, A=3}^{\rm GCF} (k) $(solid line) for $ ^3\text{He} $ with local potentials. The upper panel shows the results in a logarithmic scale, while the lower panel displays the ratio $n_{p, A=3}(k)/n_{p, A=3}^{\rm GCF} (k)$. The gray band indicates the 20\% deviation from unity.}
\label{fig:1bmd_loc_he3}
\end{figure}

\begin{figure}[htbp]
\includegraphics[width=1.0\linewidth]{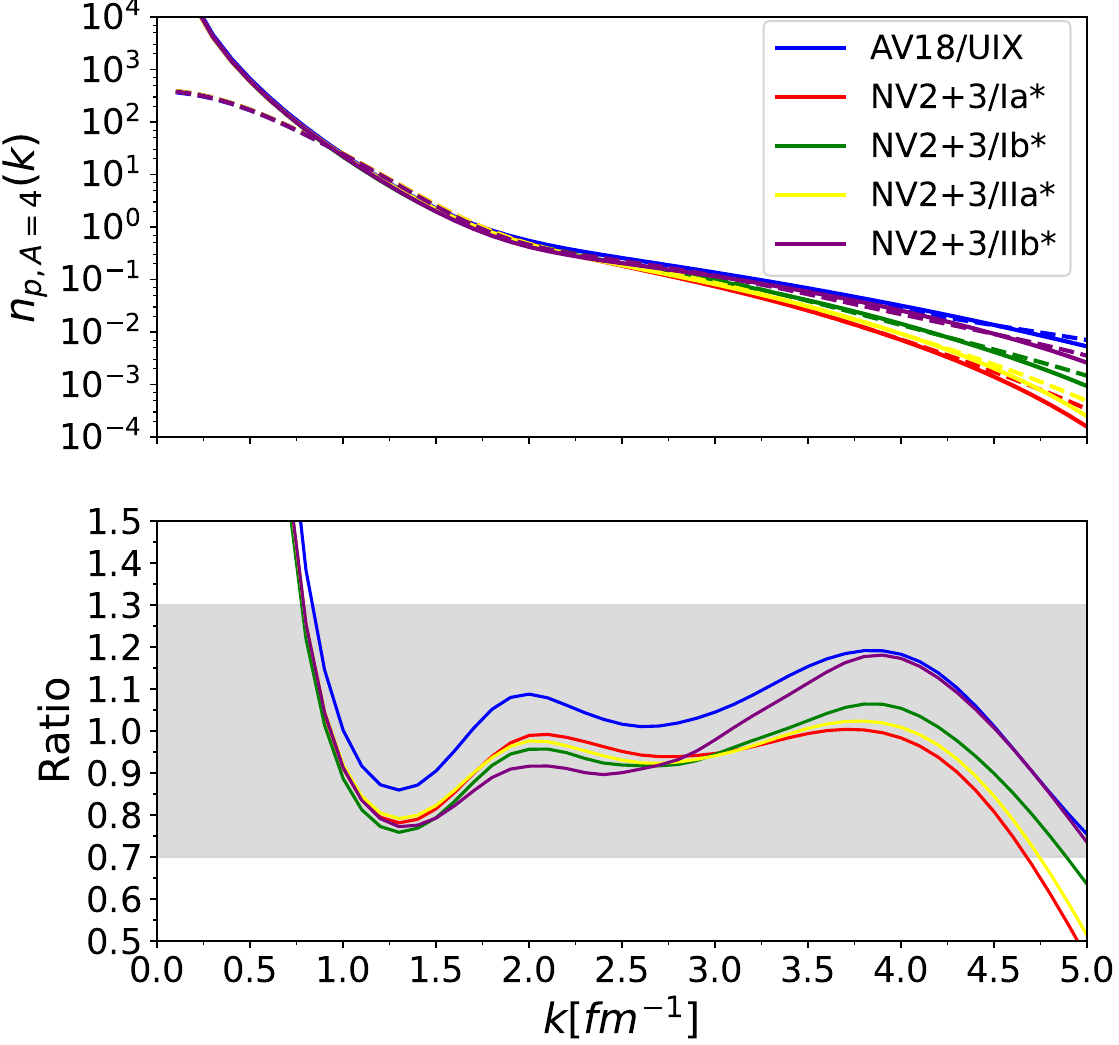}
\caption{Same as Fig.~\ref{fig:1bmd_loc_he3}, but for the $^4$He.
Here the gray band indicates the 30\% deviation from unity.}
\label{fig:1bmd_loc_he4}
\end{figure}

As a final test, we calculated the one-body momentum distribution (1BMD) for each nucleus. In the case of a proton and $^3$He as an example, the 1BMD is defined as~\cite{Marcucci:2018llz}
\begin{equation}
        n_{p,A}(k)= \frac{1}{Z} \int d{\hat{\bm k}} \int d{\bm K} \psi^\dagger(\bm{k},\bm{K}) \, P_p \psi(\bm{k},\bm{K})\ , \label{eq:1bmd-def}
\end{equation}
where $P_p$ is the proton projection operator, $\bm{k}$ the proton momentum and $\bm{K}$ the remaining pair relative momentum. Here $Z=2$.
 Within the GCF framework, this 1BMD should be given by~\cite{PhysRevC.92.054311,WEISS2018211}
\begin{eqnarray}
    n_{p,A}^{\rm GCF}(k) &= & \, 2 \tilde{C}_{pp,A}^{S=0} |\tilde{\varphi}_{pp}^{S=0}(k)|^2 \, + \,  \tilde{C}_{np,A}^{S=0} |\tilde{\varphi}_{np}^{S=0}(k)|^2  \nonumber \\
    & +  & \tilde{C}_{np,A}^{S=1} |\tilde{\varphi}_{np}^{S=1}(k)|^2\ . \label{1bmd}
\end{eqnarray}
We report here the results for $ ^3\text{He} $ and $ ^4\text{He} $, as the behavior for $ ^3\text{H} $ follows that of $ ^3\text{He} $. 
Note that in the case of $^3$H, though, Eqs.~\eqref{eq:1bmd-def} and~\eqref{1bmd} should be applied with the exchange $p\leftrightarrow n$. 
The agreement between $n_{p,A}(k)$ of Eq.~\eqref{eq:1bmd-def} and $n_{p,A}^{\rm GCF}(k)$ of Eq.~\eqref{1bmd} represents a test for the GCF formalism.
In Figs.~\ref{fig:1bmd_loc_he3} and~\ref{fig:1bmd_loc_he4} we show the results for $^3$He and $^4$He obtained with local potentials.
From inspection of these figures, we can observe that $n_{p,A}(k)$ and $n_{p,A}^{\rm GCF}(k)$remain within 20\% in the range $k\sim 1.5-4.0$ fm$^{-1}$ for both $^3$He and $^4$He.  The $^3$He results, though, are more stable and the ratios closer to 1 than in the $^4$He case.

\begin{figure}[htbp]
\includegraphics[width=1.0\linewidth]{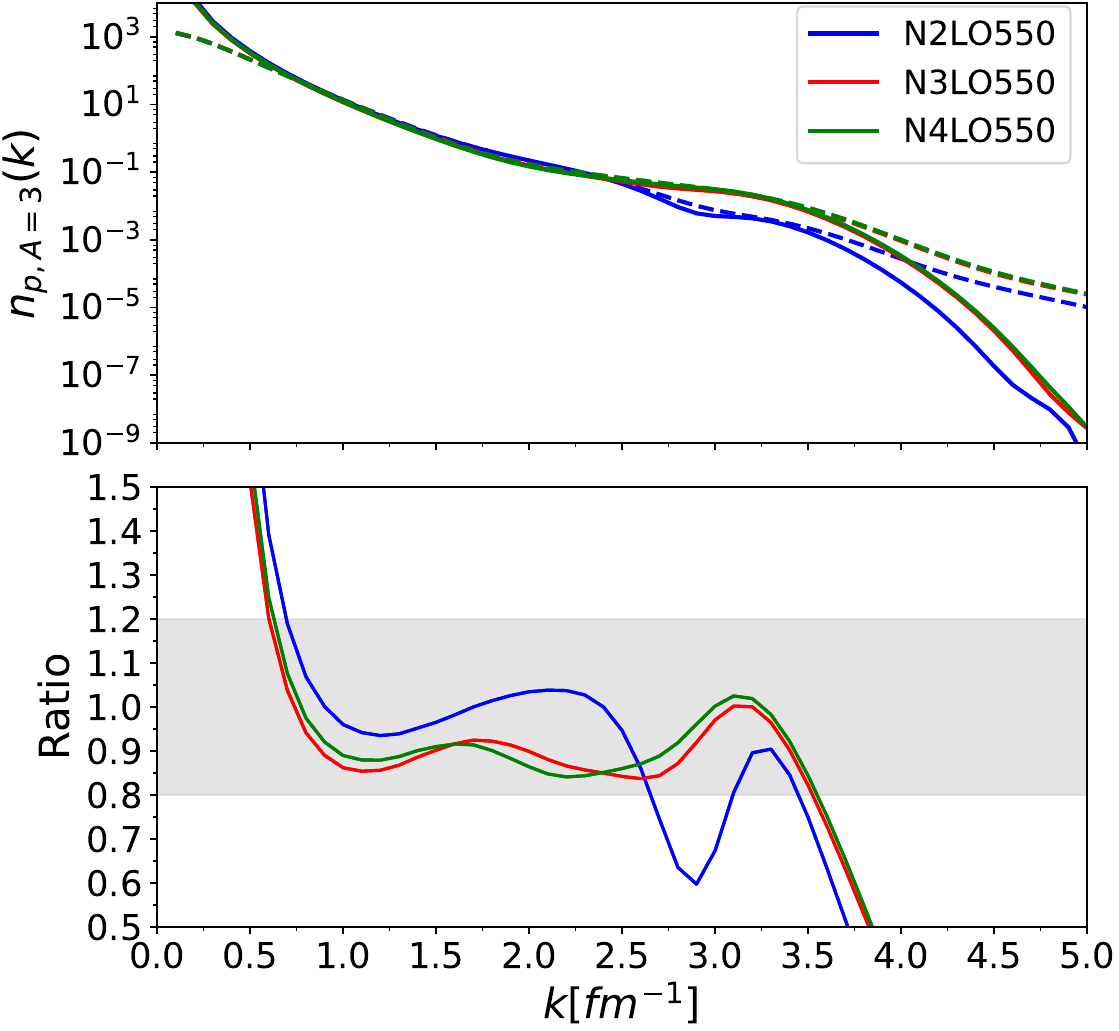}
\caption{Same as Fig.~\ref{fig:1bmd_loc_he3}, but for the N2LO, N3LO and N4LO chiral potentials with fixed cutoff $\Lambda=550$ MeV.}
\label{fig:1bmd_550_he3}
\end{figure}

\begin{figure}[htbp]
\includegraphics[width=1.0\linewidth]{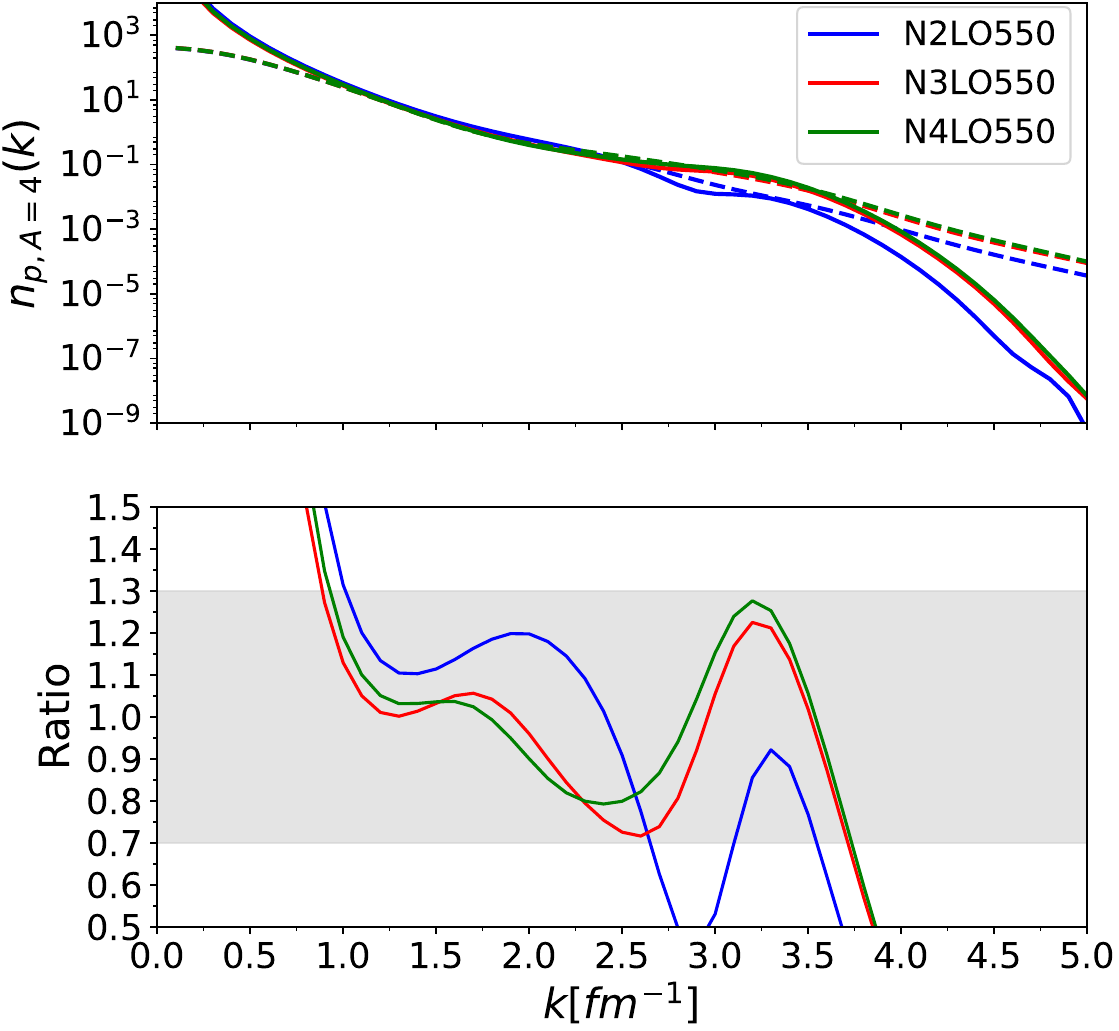}
\caption{Same as Fig.~\ref{fig:1bmd_loc_he4}, but for the N2LO, N3LO and N4LO chiral potentials with fixed cutoff $\Lambda=550$ MeV.}
          \label{fig:1bmd_550_he4}
\end{figure}

In Figs.~\ref{fig:1bmd_550_he3} and~\ref{fig:1bmd_550_he4} we report the results obtained with the chiral potentials N2LO, N3LO, and N4LO with cutoff $ \Lambda = 550 $ MeV, which is the one presenting the largest difference between $n_{p,A}(k)$ and $n_{p,A}^{\rm GCF}(k)$. Specifically, from the figures, we first observe that for large values of $ k $ (above about $ 3.5 \, \text{fm}^{-1} $), cutoff effects arise, which, as expected, invalidate any conclusion. These effects appear for larger values of $ k $ as the cutoff increases. 
As with local potentials, we also observe that the results for $ ^3\text{He} $ are of a better quality than those for $ ^4\text{He} $. 
  In particular, the ratio $n_{p,A}(k)/n_{p,A}^{\rm GCF}(k)$ in the case of $ ^3\text{He} $ ($ ^4\text{He} $) remains within $ 20\% $ ($ 30\% $) from unity in the range of $k\sim 1.0 - 3.5$ fm$^{-1}$ ($k\sim 1.0 - 3.0$ fm$^{-1}$) for the N3LO and N4LO potentials. After this range, cutoff effects emerge. The N2LO results present larger fluctuations around 1 than the results obtained with the potentials of other chiral orders, reducing the range of $k$ where the ratio is within 20\% (30\%) from unity to $k\sim 1.0 - 2.5$ fm$^{-1}$ for both $ ^3\text{He} $ and $ ^4\text{He} $. This is true for any cutoff value, although we should remark that $\Lambda=550$ MeV is the worst case.

  The conclusion of this test, performed for the first time using both
    local and non-local interactions, is that the GCF predictions are satisfied
    at the 20\% or 30\% level for $ ^3\text{He} $ and $ ^4\text{He} $,
    respectively, and therefore they present
    some tension, especially in the case of non-local potentials.

\subsection{2BMDs and 2BDFs vs.\ universal functions}
\label{subsec:2b-uf}

In this subsection we report the results obtained for the ratios of the 2BDFs and 2BMDs to the universal functions $\varphi_{N_1 N_2}^S(r)$ or their FT $\tilde\varphi_{N_1 N_2}^S(r)$. These ratios are needed to extract the contact coefficients as defined in Eqs.~\eqref{eq:cpp0_old_r}--\eqref{eq:cnp0_old_r} and~\eqref{eq:cpp0_old_k}--\eqref{eq:cnp0_old_k}, with the assumption $L=0$, and in Eqs.~\eqref{eq:cpp0_r}--\eqref{eq:cnp0_r} and~\eqref{eq:cpp0_k}--\eqref{eq:cnp0_k} without this assumption. 
The considered $N_1 N_2$ channels are $ nn $ (or $ pp $) with $ S=0 $, $ np $ with $ S=0 $, and $ np $ with $ S=1 $.

The ratios have been calculated for all local and non-local potentials listed at the end of Sec.~\ref{subsec:2BDFMD} for the $ ^3\mathrm{H} $, $ ^3\mathrm{He} $, and $ ^4\mathrm{He} $ nuclei. 
Given the large number of cases we have analyzed, here we present only a subset for $^3$He (and later for $^4$He), since the results for $ ^3\mathrm{H} $ and $ ^3\mathrm{He} $ exhibit analogous trends.
In particular, we focus on the results obtained with the local NV2+3/Ia* and NV2+3/Ib* Norfolk potentials, because the NV2+3/IIa* and NV2+3/IIb* results are very similar. Regarding the for $A=3$ obtained with 
non-local potentials, we include those at various chiral orders with cutoffs $ \Lambda = 450$ MeV and $500 $ MeV, as the results obtained with $ \Lambda = 550 $ MeV exhibit a behavior analogous to the $ \Lambda = 500 $ MeV ones. The cutoff dependence is studied in the ratios of 2BDFs at
N2LO and N3LO, as the behavior at N4LO closely resembles that at N3LO.

\begin{figure*}[htbp]
\centering
\includegraphics[width=\textwidth]{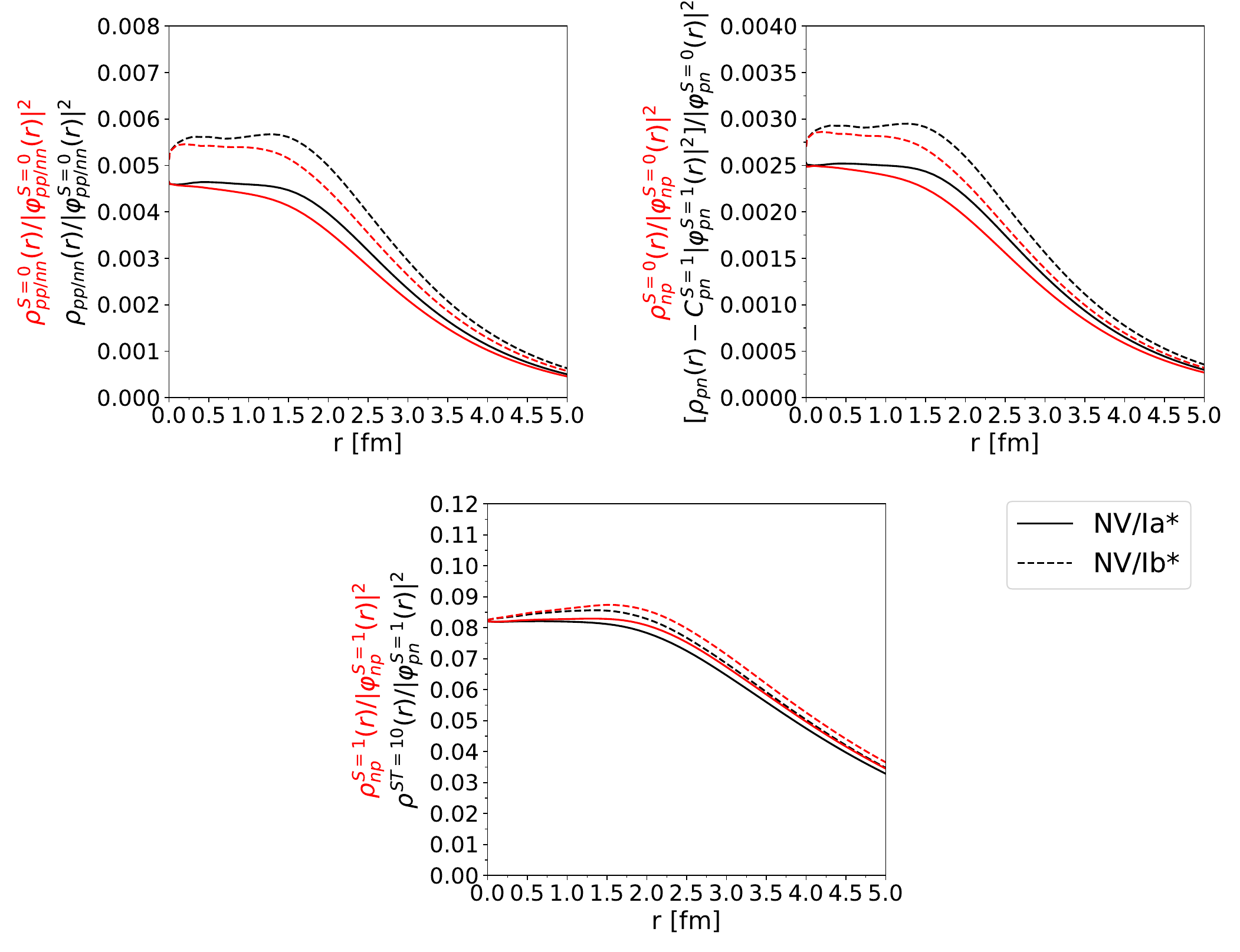}
\caption{Ratios between the 2BDFs and the universal functions, with (black lines) or without (red lines) the $L=0$ assumption for the $^3$He nucleus, calculated using the Norfolk potentials  NV2+3/Ia* and NV2+3/Ib*.}
\label{fig:2bdf_ratio_loc}
\end{figure*}
\begin{figure*}[htbp]
\centering
\includegraphics[width=\textwidth]{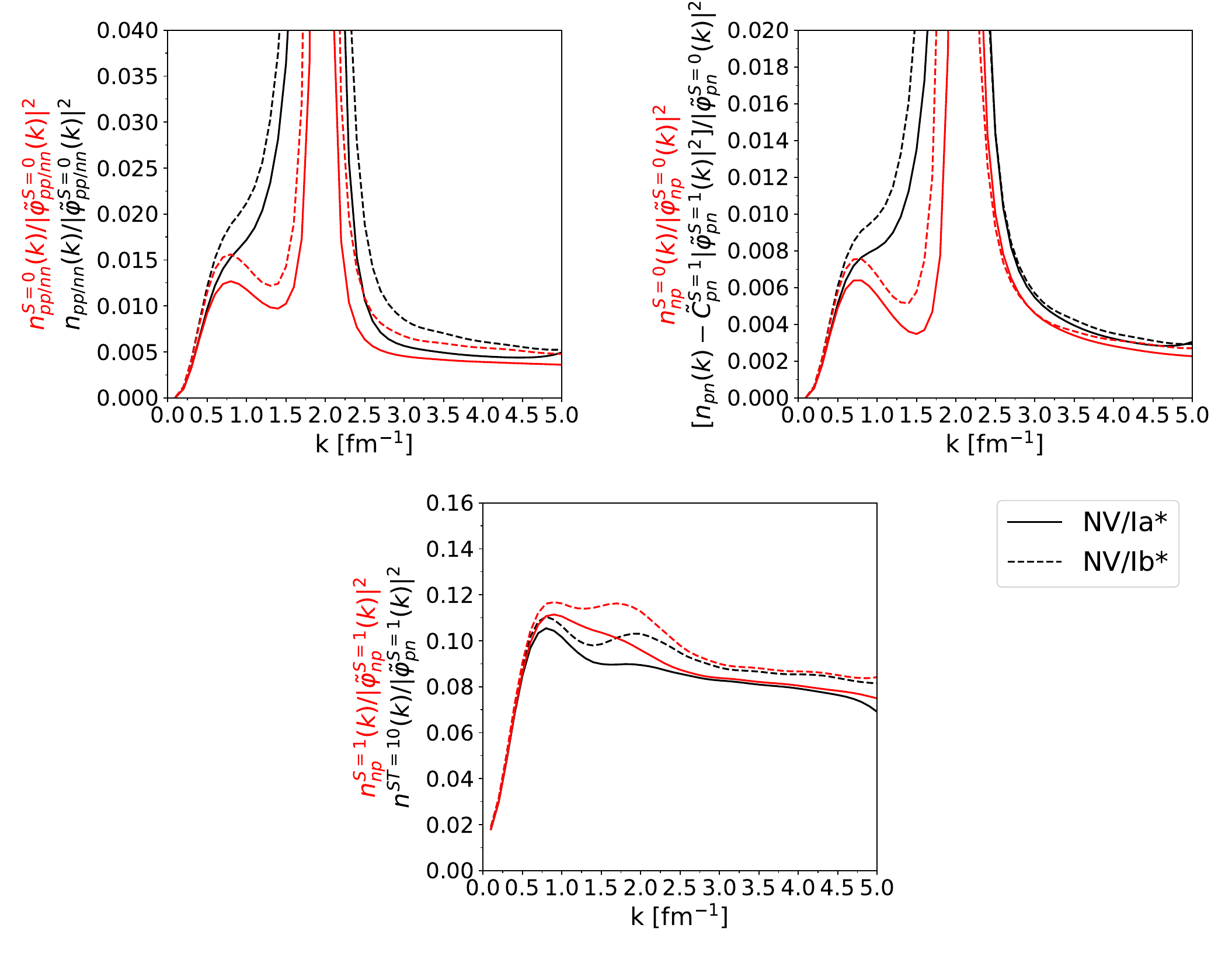}
\caption{Same as Fig.~\ref{fig:2bdf_ratio_loc}, but for the 2BMDs and the FT of the universal function.}
\label{fig:2bmd_ratio_loc}
\end{figure*}
We present in Figs.~\ref{fig:2bdf_ratio_loc} and~\ref{fig:2bmd_ratio_loc} the ratios between the 2BDFs or 2BMDs and the universal functions, or their FT, with and without the $L=0$ assumption (see Eqs.~\eqref{eq:cpp0_old_k}--\eqref{eq:cnp0_r}), calculated using the local potentials NV2+3/Ia* and NV2+3/Ib* for the $^3$He nucleus. By inspection of the figures we can see that plateaus are observed for large values of $ k $ and small values of $ r $, corresponding to regions dominated by SRCs.
Specifically, in Fig.~\ref{fig:2bmd_ratio_loc}, the plateau region occurs at very high $ k $. For this reason, we choose to extract the contact coefficients in the range $ 3.5-4.5 \, \text{fm}^{-1} $. On the other hand, in Fig.~\ref{fig:2bdf_ratio_loc}, the plateau appears for small values of $ r $. Therefore, we have decided to extract the coefficients in the range $ 0-1 \, \text{fm} $. 
The choice of these ranges for $k$ and $r$ is consistent with Ref.~\cite{Cruz-Torres:2019fum}.
\begin{figure*}[htbp]
\centering
\includegraphics[width=\linewidth]{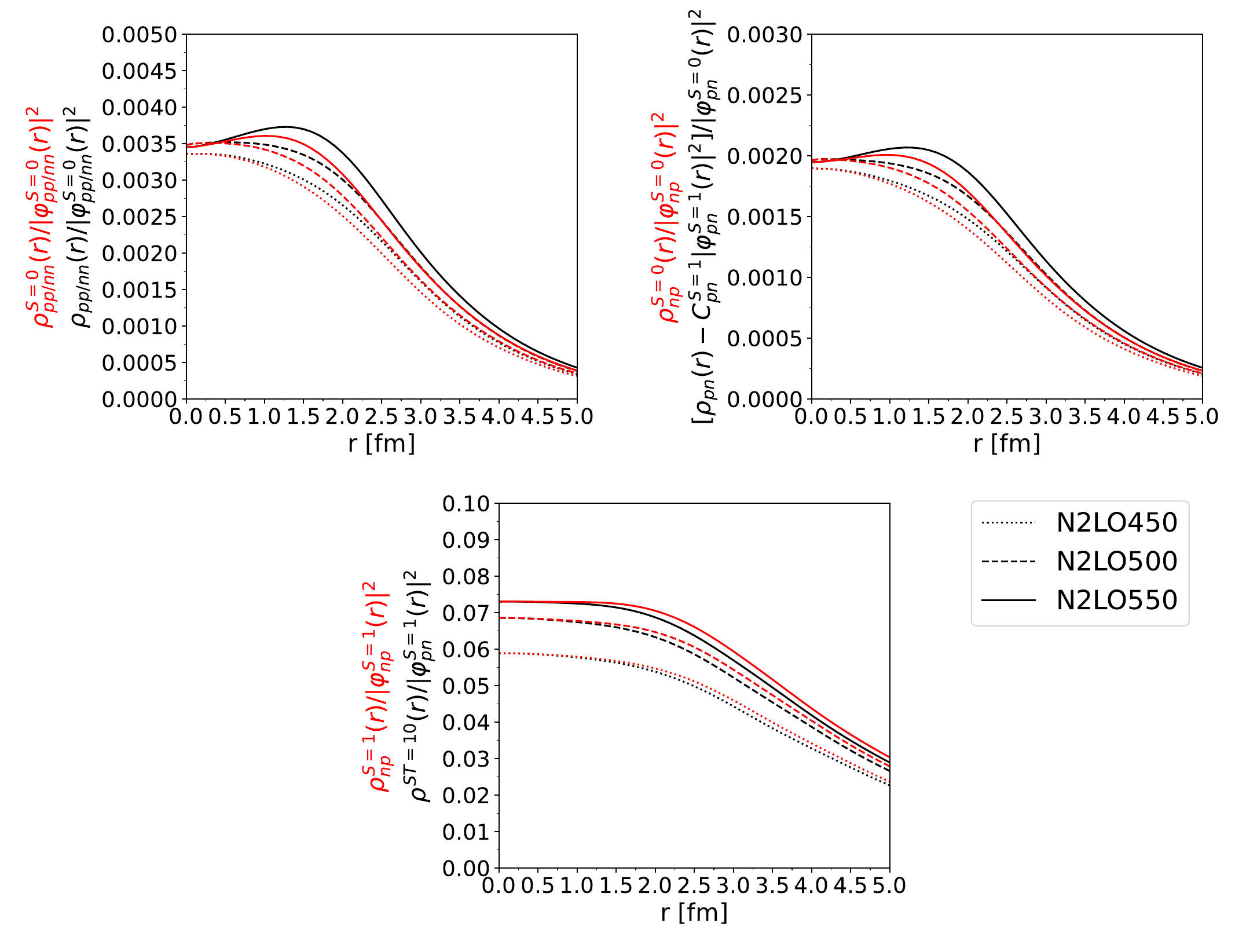}
\caption{Same as Fig.~\ref{fig:2bdf_ratio_loc}, but for the non-local potentials N2LO450, N2LO500 and N2LO550.}
\label{fig:2bdf_ratio_nonloc_n2lo}
\end{figure*}
\begin{figure*}[htbp]
\centering
\includegraphics[width=\linewidth]{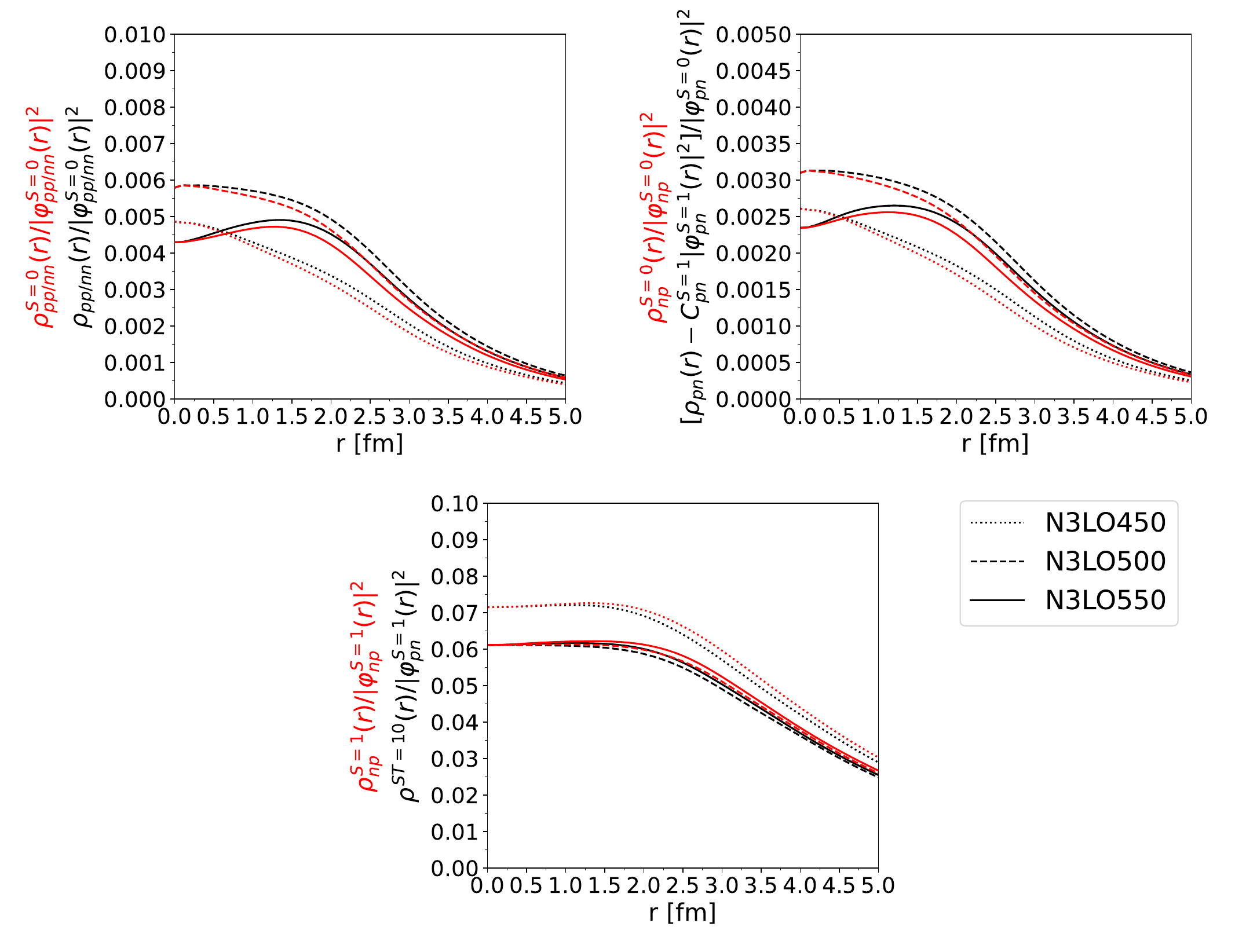}
\caption{Same as Fig.~\ref{fig:2bdf_ratio_nonloc_n2lo}, but for the non-local potentials N3LO450, N3LO500 and N3LO550.}
\label{fig:2bdf_ratio_nonloc_n3lo}
\end{figure*}
\begin{figure*}[htbp]
\centering
\includegraphics[width=\linewidth]{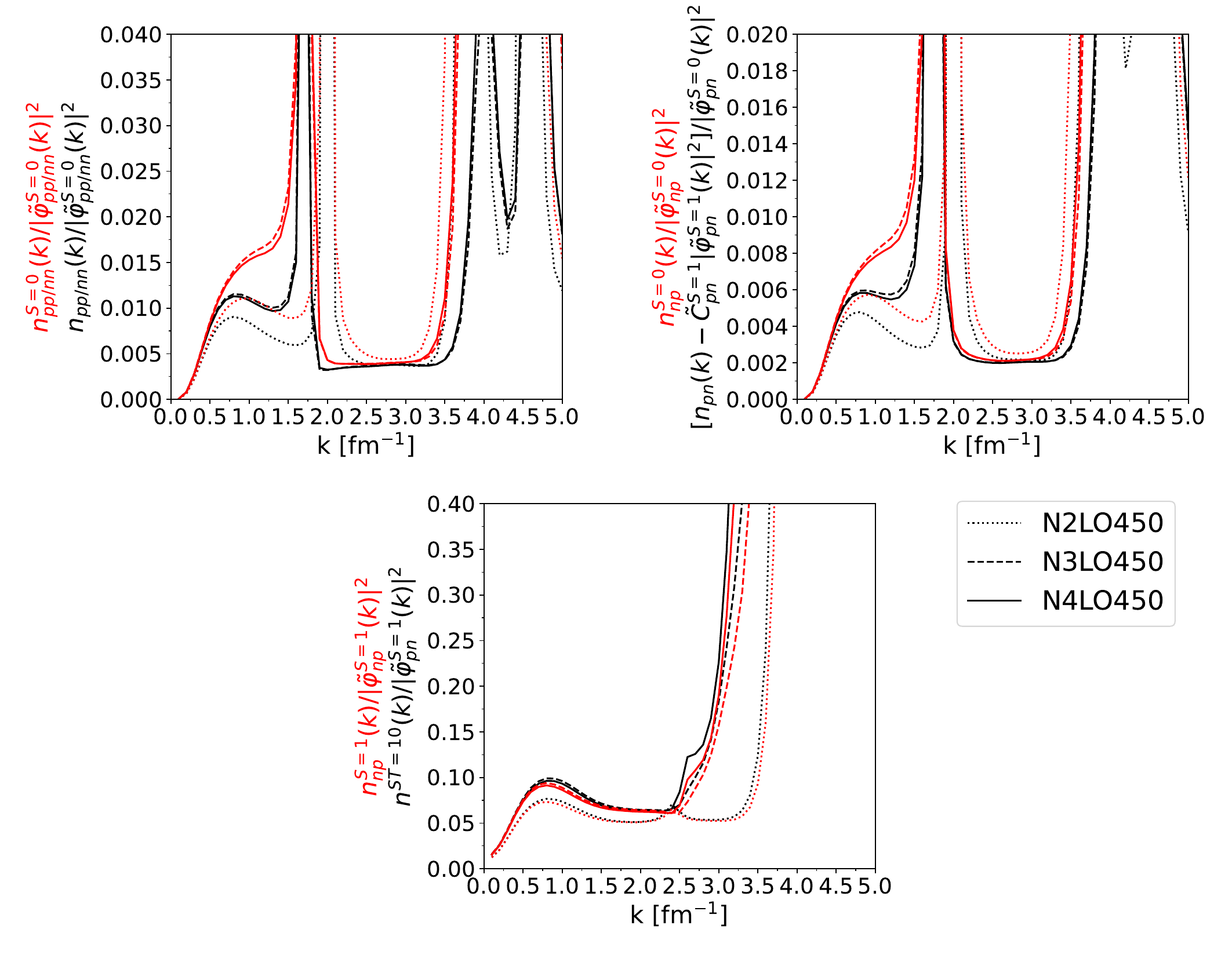}
\caption{Same as Fig.~\ref{fig:2bmd_ratio_loc}, but for the non-local potentials N2LO450, N3LO450 and N4LO450.}
\label{fig:2bmd_ratio_nonloc_450}
\end{figure*}
\begin{figure*}[htbp]
\centering
\includegraphics[width=\linewidth]{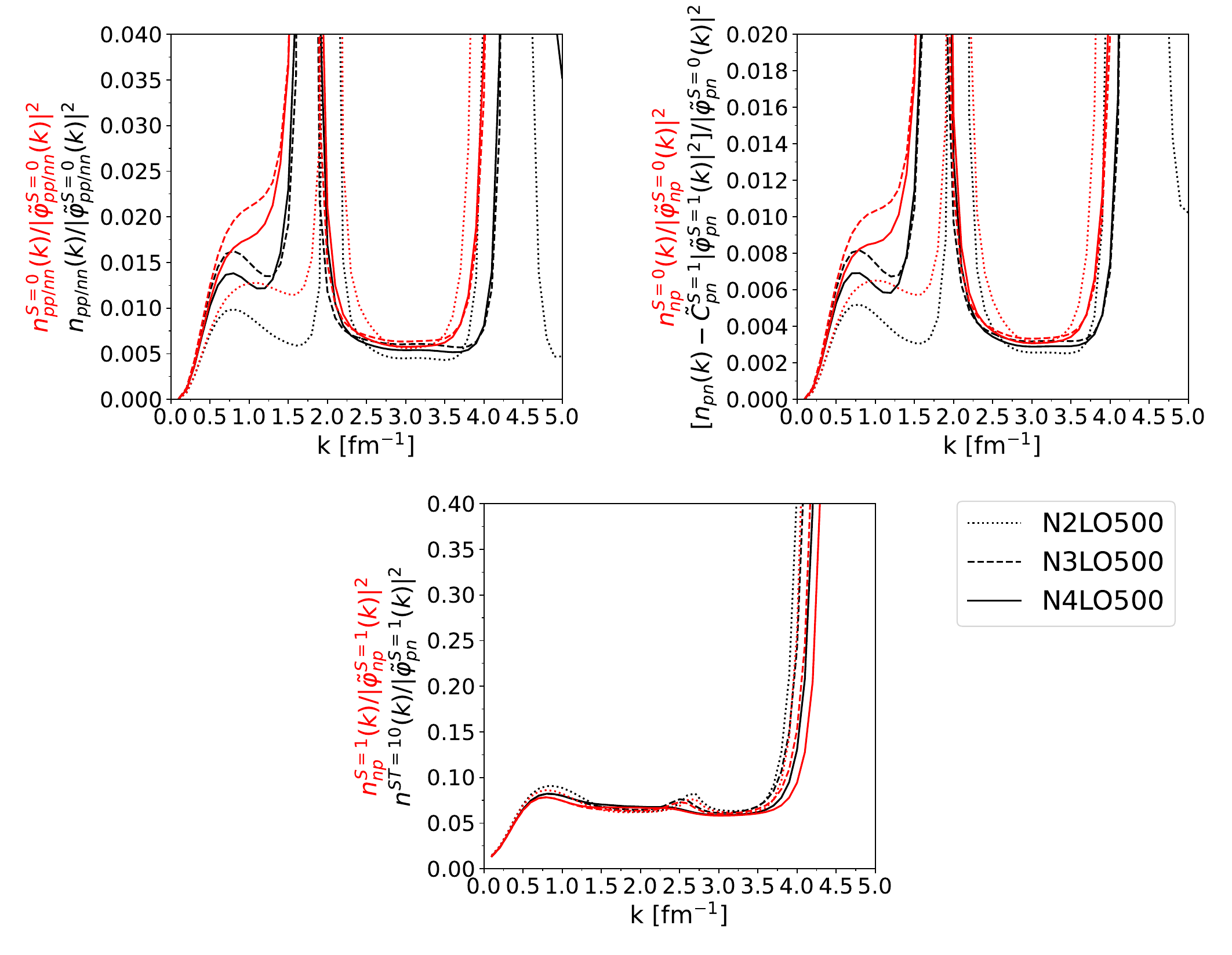}
\caption{Same as Fig.~\ref{fig:2bmd_ratio_nonloc_450}, but for the non-local potentials N2LO500, N3LO500 and N4LO500.}
\label{fig:2bmd_ratio_nonloc_500}
\end{figure*}

We present in Figs.~\ref{fig:2bdf_ratio_nonloc_n2lo}, \ref{fig:2bdf_ratio_nonloc_n3lo}, \ref{fig:2bmd_ratio_nonloc_450} and~\ref{fig:2bmd_ratio_nonloc_500}
the ratios between the 2BDFs or 2BMDs and the universal functions, or their FT, with and without the $L=0$ assumption calculated using the non-local potentials for the $^3$He nucleus. By inspection of the Figs.~\ref{fig:2bdf_ratio_nonloc_n2lo} and~\ref{fig:2bdf_ratio_nonloc_n3lo} we can see that a clear plateau appears for $ r\to 0 $. Consequently, the contact coefficients are extracted in the range $ 0 \text{--} 1 \, \mathrm{fm} $. However, differences arise depending on the spin channel and the cutoff $\Lambda$. Specifically, the plateau in the $ S = 1 $ channel is significantly larger and more pronounced compared to the $ S = 0 $ channels for all potentials.
Additionally, in the $ S = 0 $ channels, differences depending on the cutoff are evident at the same chiral order. In particular, for $\Lambda = 500$ MeV, the plateau is the most pronounced, while for $\Lambda = 450$ MeV, the curve exhibits an increasing trend as $ r \to 0 $. Conversely, for $\Lambda = 550$ MeV, a decreasing trend is observed as $ r \to 0 $. These differences are more evident for N3LO and N4LO (the latter, which exhibits behavior similar to N3LO, is not shown) compared to N2LO.
Despite these cutoff effects at small $ r $, it can be observed that all curves between $ 0 $ and $ 1 \, \mathrm{fm} $ are relatively flat.

By inspection of Figs.~\ref{fig:2bmd_ratio_nonloc_450} and~\ref{fig:2bmd_ratio_nonloc_500} 
we can see that for non-local potentials, the plateau region is found for slightly smaller values of $ k $ compared with the local potentials. The high-momentum divergences are very likely due to cutoff effects. As a consequence, the corresponding range of $ k $ cannot be considered. Additionally, it can be observed that the plateau region depends on the spin-channel and the cutoff $\Lambda$. Specifically, coefficients for $ S = 0 $ and $ S = 1 $ are extracted in different regions, as the plateau appears at different values of $ k $ even for the same potential.
Finally, it can also be observed that while for $ S = 0 $ the graphs for $ \Lambda = 450 $ MeV and $ \Lambda = 500 $ MeV are similar, for $ S = 1 $ substantial differences are visible. More in detail, in the case of $ S = 1 $, for the N3LO450 and N4LO450 potentials, only one plateau is observed (around $ 1.5-2 \, \mathrm{fm}^{-1} $), while for the N2LO450 potential, two plateaus are present: the one discussed above and another corresponding to larger $ k $ values. To ensure uniformity, we have chosen to extract the coefficients at the plateau corresponding to the smaller values of $ k $ for all the three potentials with $ \Lambda = 450 $ MeV.
On the other side, for the $ \Lambda = 500 $ MeV potentials, two plateaus are observed for all three orders (see Fig.~\ref{fig:2bmd_ratio_nonloc_500} ): one analogous to the plateau seen for $ \Lambda = 450 $ MeV and another at higher values of $ k $. Consequently, we decided to extract two values of $ C_{np}^{S=1} $, one for each plateau, and calculate the final coefficient by averaging these two values. The same approach was applied for $ \Lambda = 550 $ MeV.

Inspecting Figs.~\ref{fig:2bdf_ratio_loc}--\ref{fig:2bmd_ratio_nonloc_500}, we can also observe that the differences between the calculations with and without the $ L=0 $ approximation are evident, and in some cases in the $L=0$ approximation is very difficult to identify a plateau. These pose severe questions to the predictions of the GCF for non-local potentials, when the $L=0$ approximation is used. 
Nevertheless, differences arise depending on the spin channel and the potential. In general, for $ S=1 $, for both 2BDFs and 2BMDs, the differences between the curves obtained with and without the $L=0$ assumption are much smaller compared to the $ S=0 $ cases for all potentials, both local and non-local.
Remarkable is the case of the 2BMDs ratios for the $S=0$ channels, calculated with the N2LO potentials. As it can be seen from Figs.~\ref{fig:2bmd_ratio_nonloc_450} and~\ref{fig:2bmd_ratio_nonloc_500}, both the $\Lambda=450$ MeV and $\Lambda=500$ MeV potentials exhibit larger differences between the calculations with and without the $ L=0 $ approximation compared to the other chiral orders. {In fact, especially for the N2LO450 potential, the plateaus in the ratios calculated with the approximation $L=0$ are barely visible.
In the case of the local potentials results in Figs.~\ref{fig:2bdf_ratio_loc} and~\ref{fig:2bmd_ratio_loc}, differences are also observed depending on the specific potential. For example, for the $S=0$ channels, the results for the 2BDFs and 2BMDs ratios obtained with the NV2+3/Ib* potential show greater differences between the two calculations with and without the $L=0$ approximation compared to the results obtained with the NV2+3/Ia* potential. 
Additionally, we observe that for the 2BMDs, the plateaus are better defined when the $ L=0 $ approximation is not applied. This holds true for both local and non-local potentials, further reinforcing the need to go beyond this approximation to achieve more precise and stable extractions of the contact coefficients.

We have also analyzed the contribution of TNIs in the calculation without  the $ L = 0 $ approximation. Specifically, this was achieved by examining the ratios  to the universal functions and their FT of the 2BDFs and 2BMDs, respectively, calculated from the $^3$He wave functions obtained with and without TNIs.
As illustrative results, we show in Figs.~\ref{fig:2bdf_ratio_loc_noTNI} and~\ref{fig:2bmd_ratio_loc_noTNI} these ratios obtained with the Ia* and Ib* Norfolk potentials (here labelled NV/Ia* and NV/Ib*), with and without TNIs. As we can see by inspection of the figures, the calculations with and without TNIs are quite close, even if the NV/Ib* case presents larger differences than the NV/Ia* one.
\begin{figure*}[htbp]
\centering
\includegraphics[width=\linewidth]{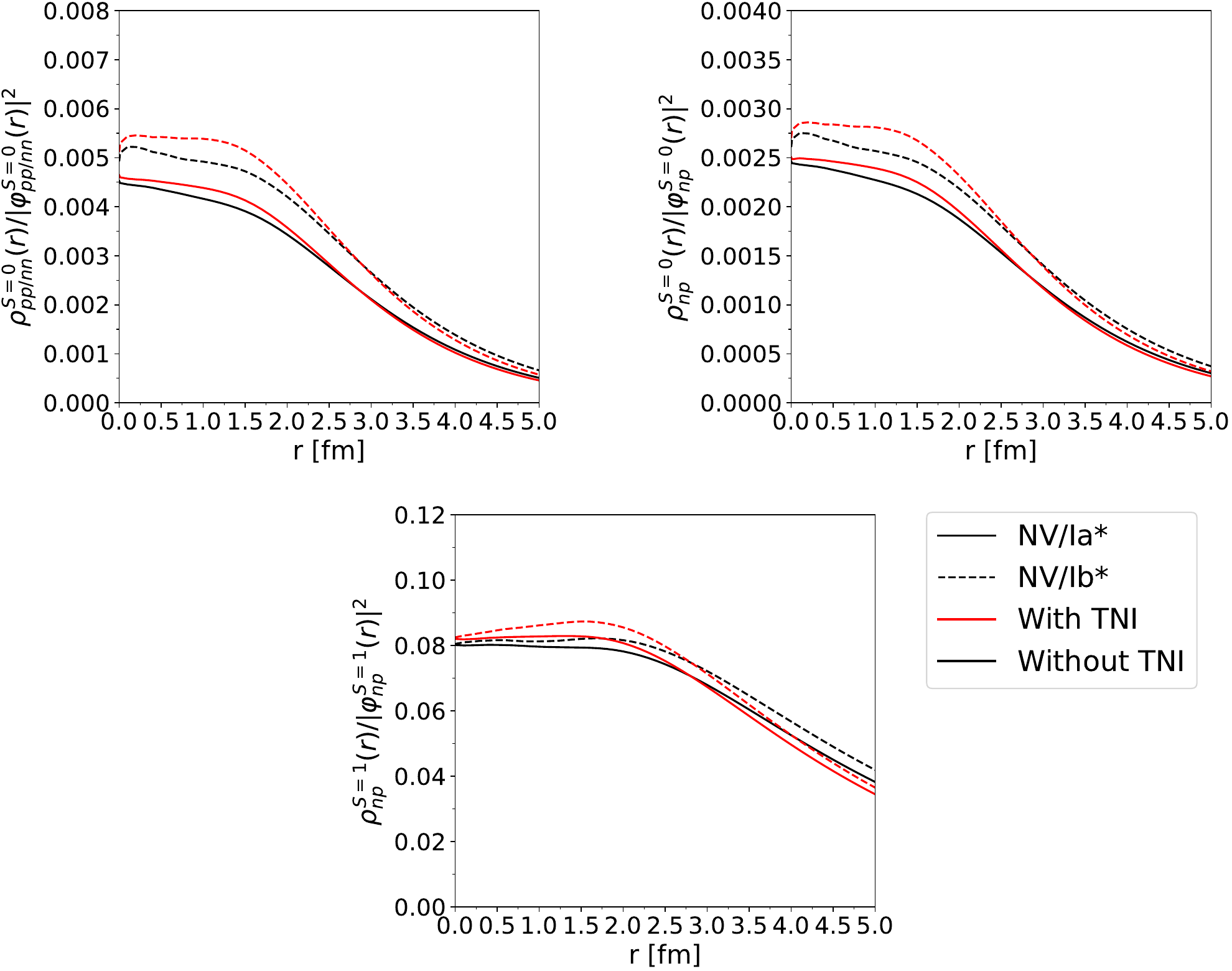}
\caption{Ratios between the 2BDFs and the universal functions, with (red lines) or without (black lines) the TNIs for the $^3$He nucleus. The calculations are performed without the $L=0$ assumption, using the Norfolk potentials NV/Ia* and NV/Ib*.}
\label{fig:2bdf_ratio_loc_noTNI}
\end{figure*}
\begin{figure*}[htbp]
\centering
\includegraphics[width=\linewidth]{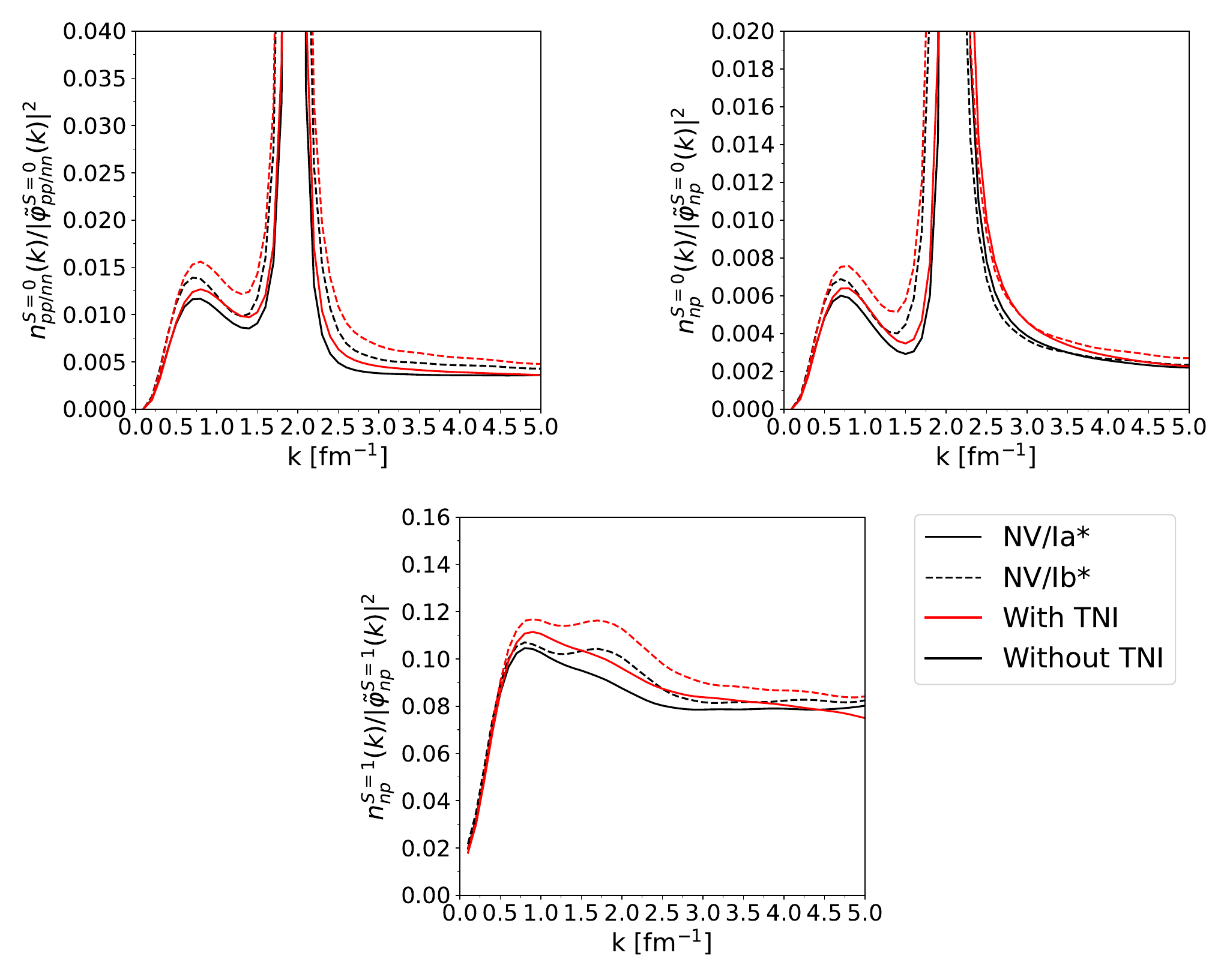}
\caption{Same as Fig.~\ref{fig:2bdf_ratio_loc_noTNI}, but for the ratios of the 2BMDs and the FT of universal functions.}
\label{fig:2bmd_ratio_loc_noTNI}
\end{figure*}
\begin{figure*}[htbp]
\centering
\includegraphics[width=\linewidth]{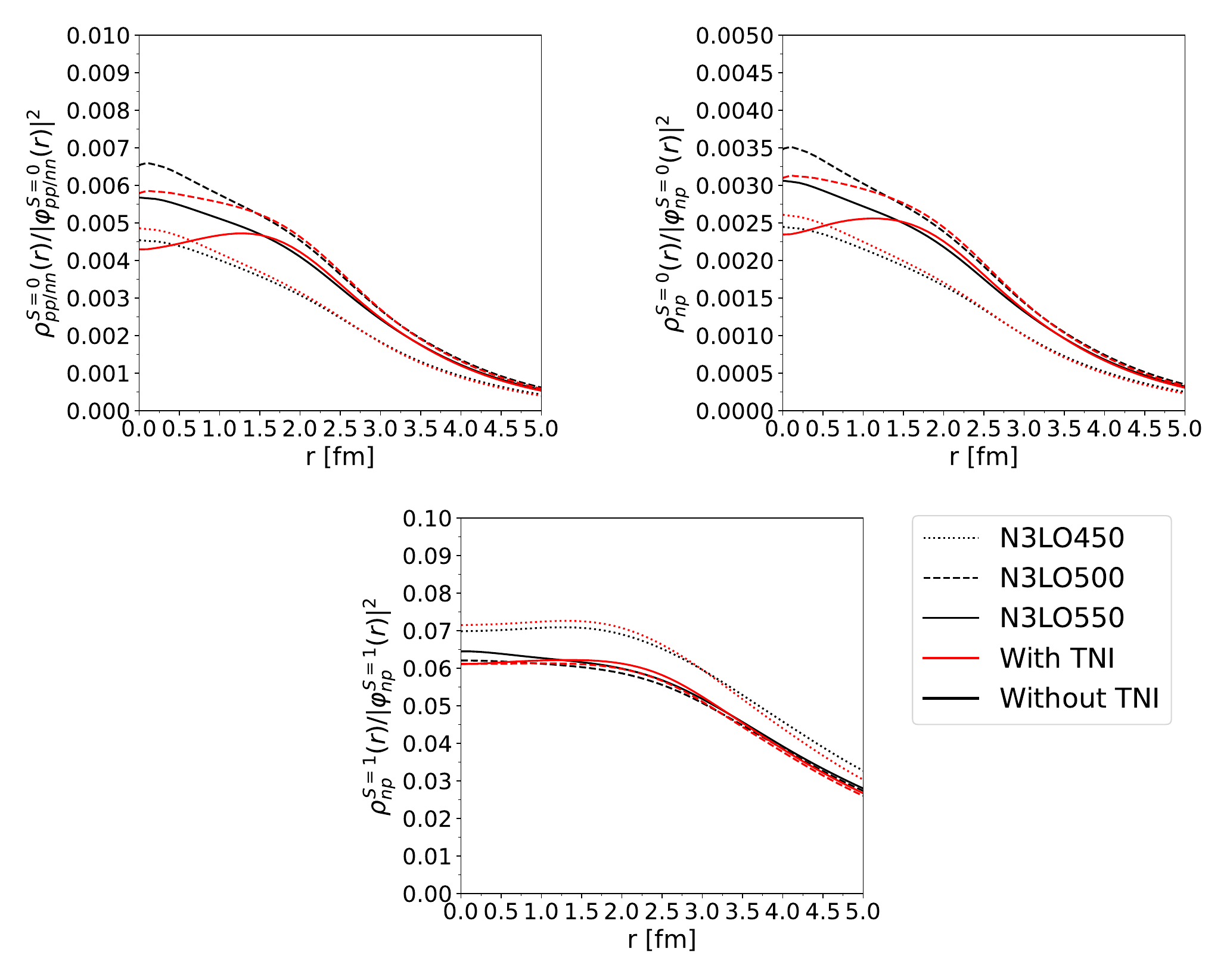}
\caption{Same as Fig.~\ref{fig:2bdf_ratio_loc_noTNI}, but for the non-local potentials with fixed chiral order (N3LO) and various cutoff values $\Lambda=450, 500, 550$ MeV.}
\label{fig:2bdf_ratio_n3lo_noTNI}
\end{figure*}
\begin{figure*}[htbp]
\centering
\includegraphics[width=\linewidth]{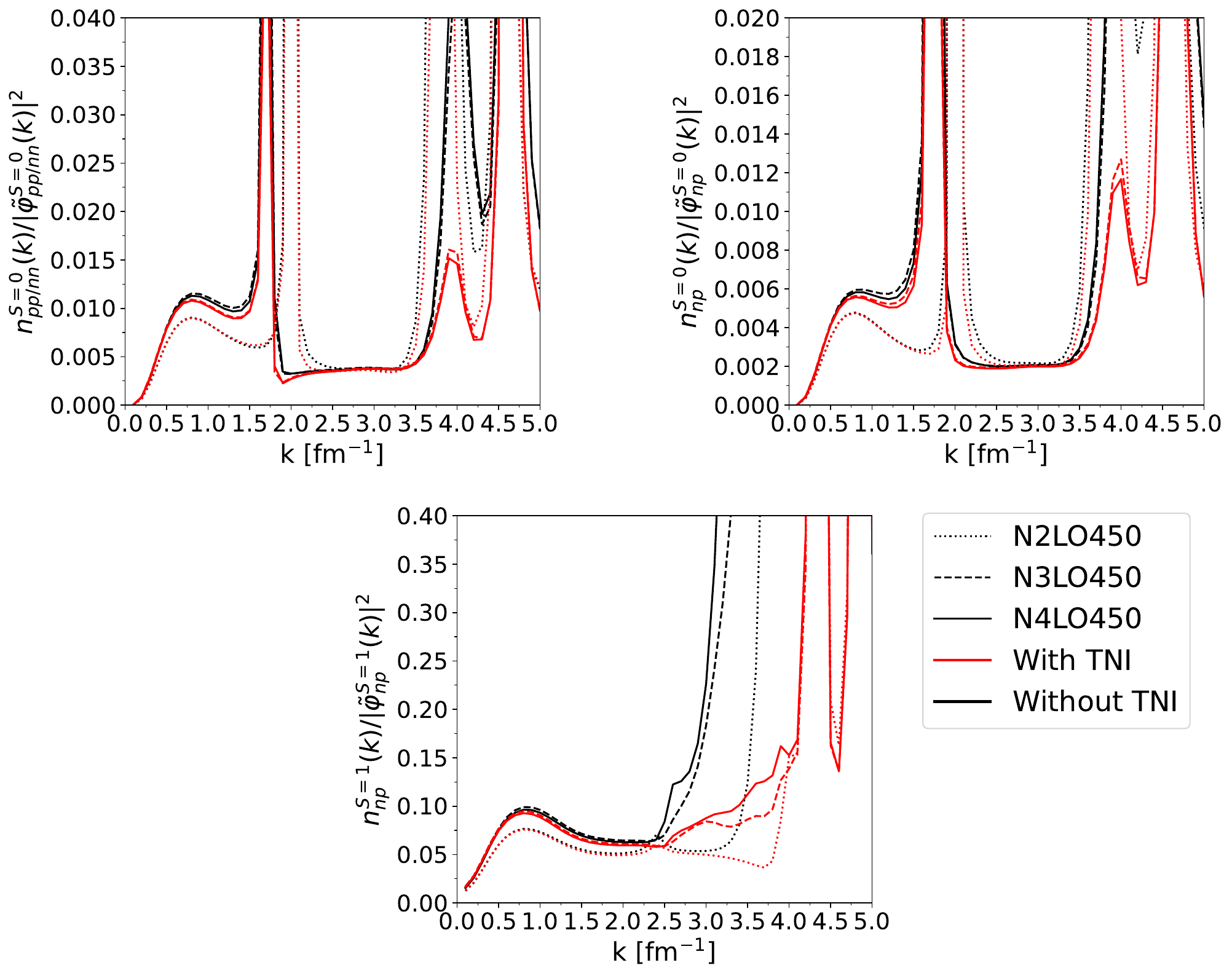}
\caption{Same as Fig.~\ref{fig:2bdf_ratio_loc_noTNI}, but for the non-local potentials with fixed cutoff value $\Lambda=450$ MeV, at various chiral order, N2LO, N3LO, and N4LO.}
\label{fig:2bmd_ratio_450_noTNI}
\end{figure*}
We present in Figs.~\ref{fig:2bdf_ratio_n3lo_noTNI} and~\ref{fig:2bmd_ratio_450_noTNI}
the results for the non-local potentials. In Fig.~\ref{fig:2bdf_ratio_n3lo_noTNI} we report the results obtained  varying the cutoff $\Lambda$ but with fixed chiral order (N3LO), because the results at N2LO or N4LO exhibit similar behaviour. As we can see from the figure, the largest differences between the calculation with and without TNIs is present in the $S=0$ channels, especially for $\Lambda=550$ MeV.
Notably, in this channel, the N2LO results exhibits smaller discrepancies compared to N3LO and N4LO results. Furthermore, we see that TNI contributions increase as the cutoff grows. For instance, while small differences are observed at $ r \to 0 $ for the N3LO450 potential, much larger differences are observed for the N3LO550 potential in the $ S = 0 $ channel.
In the dominant $ S = 1 $ channel, instead, these differences remain small, and the cutoff dependence is very mild. 
In Fig.~\ref{fig:2bmd_ratio_450_noTNI}, we report the results obtained with cutoff value of $\Lambda=450$ MeV, since this is the case which presents the largest differences between the results obtained with and without TNIs. In particular, by inspection of the figure, we can see that
in the $S=0$ channels, the N2LO results exhibit the largest differences. This is true also for $ \Lambda = 500 $ MeV and $\Lambda=550$ MeV. Moreover, we have observed that as the cutoff increases, the contribution from TNIs grows.

We move now the $^4$He results. In general, these are very similar to the $^3$He ones. Therefore, all the same conclusions hold, except for the 2BMDs ratios for the non-local chiral potentials at N2LO, for any cutoff value. As it can be seen from Figs.~\ref{fig:2bmd_ratio_nonloc_500_4he}, where the representative cutoff value of $\Lambda=500$ MeV is used, 
the discrepancies between the results with and without the $ L= 0 $ approximation in the $S=0$ channel are more pronounced compared to $ ^3\text{He} $. Specifically, the plateau obtained with the $ L = 0 $ approximation is much harder to identify, and its value is significantly higher than that obtained without the approximation. 
\begin{figure*}[htbp]
\centering
\includegraphics[width=\linewidth]{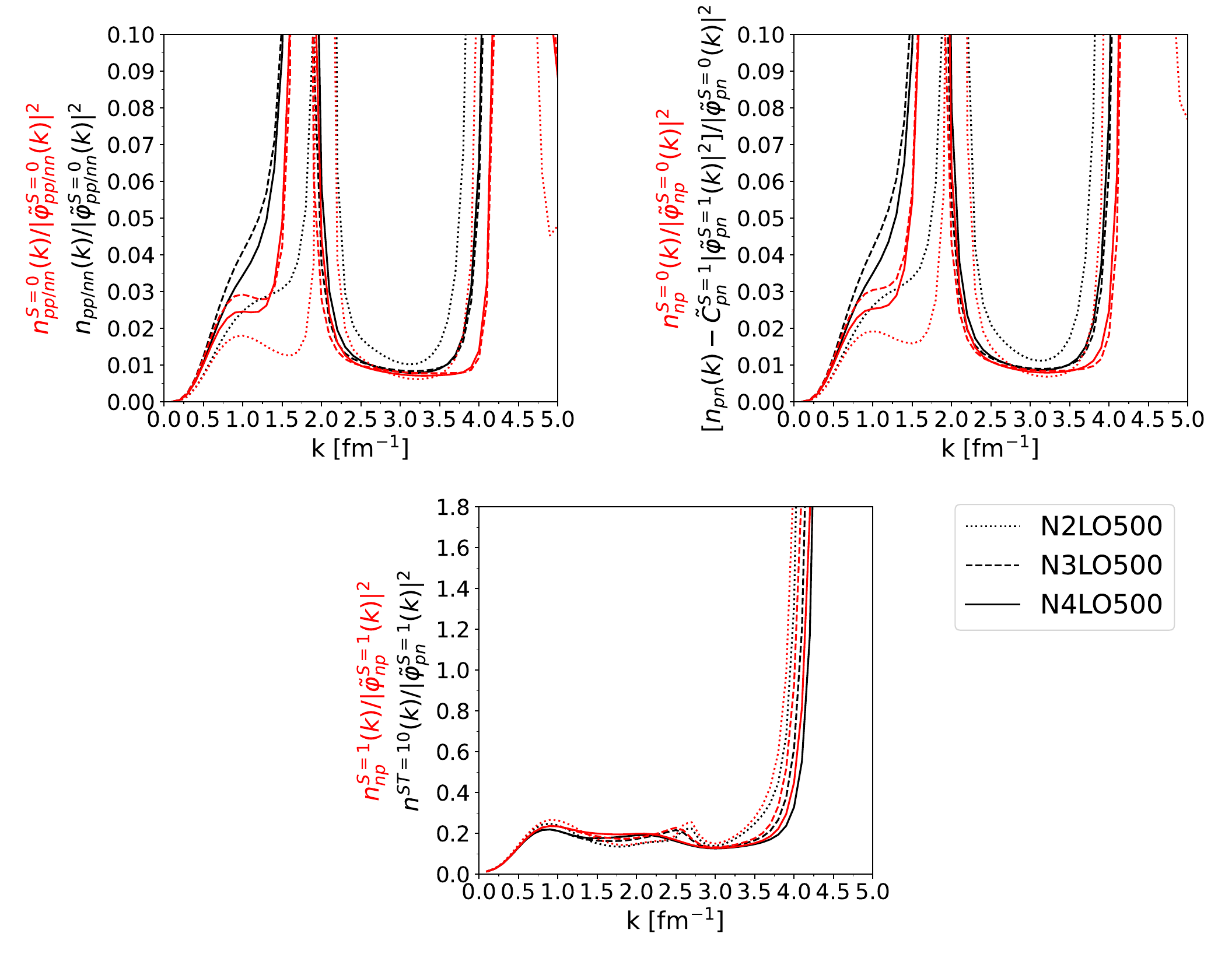}
\caption{Same as Fig.~\ref{fig:2bmd_ratio_nonloc_500}, but for the $^4$He nucleus.}
\label{fig:2bmd_ratio_nonloc_500_4he}
\end{figure*}
\begin{figure*}[htbp]
\centering
\includegraphics[width=\linewidth]{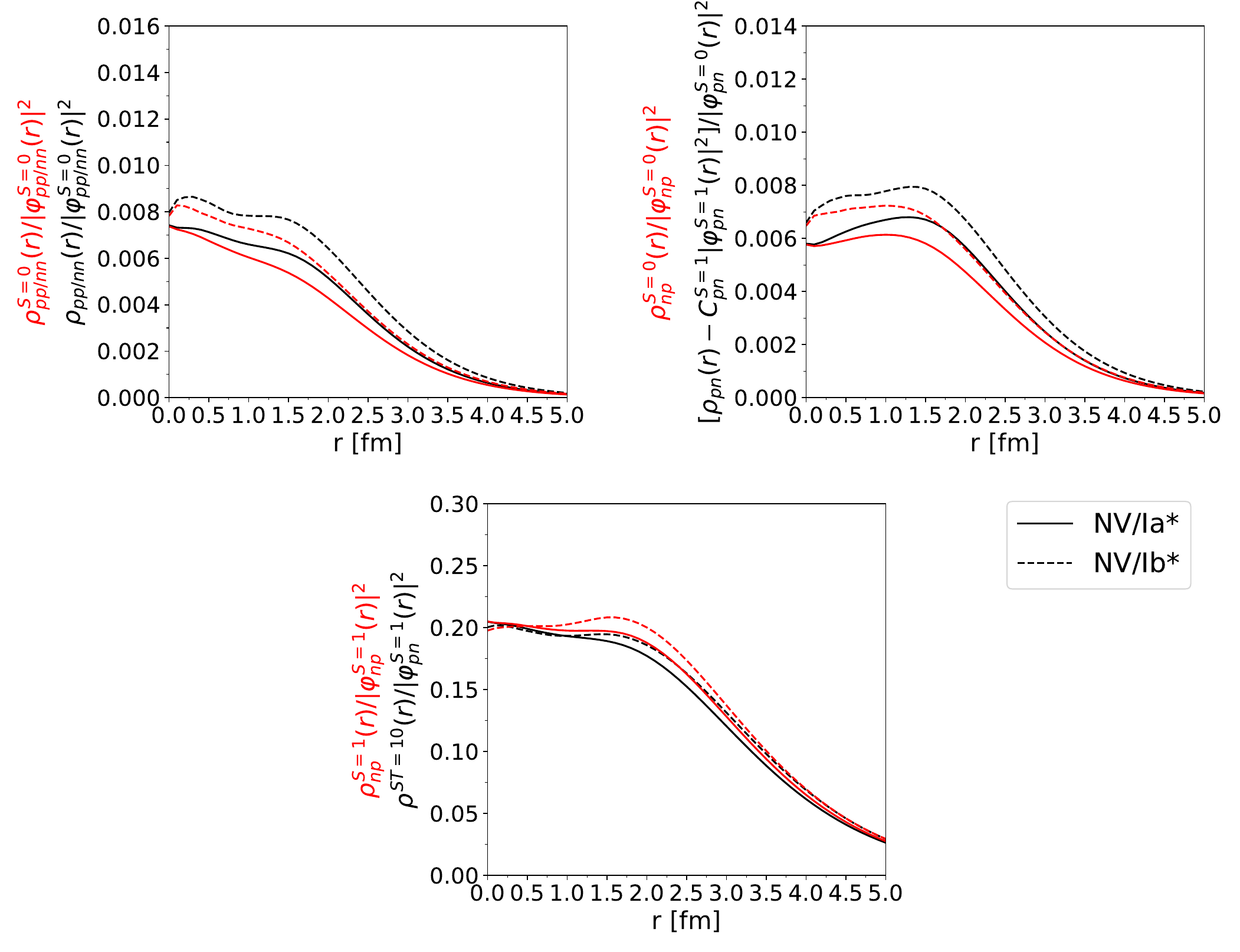}
\caption{Same as Fig.~\ref{fig:2bdf_ratio_loc}, but for the $^4$He nucleus.}
\label{fig:2bdf_ratio_loc_4he}
\end{figure*}
\begin{figure*}[htbp]
\centering
\includegraphics[width=\linewidth]{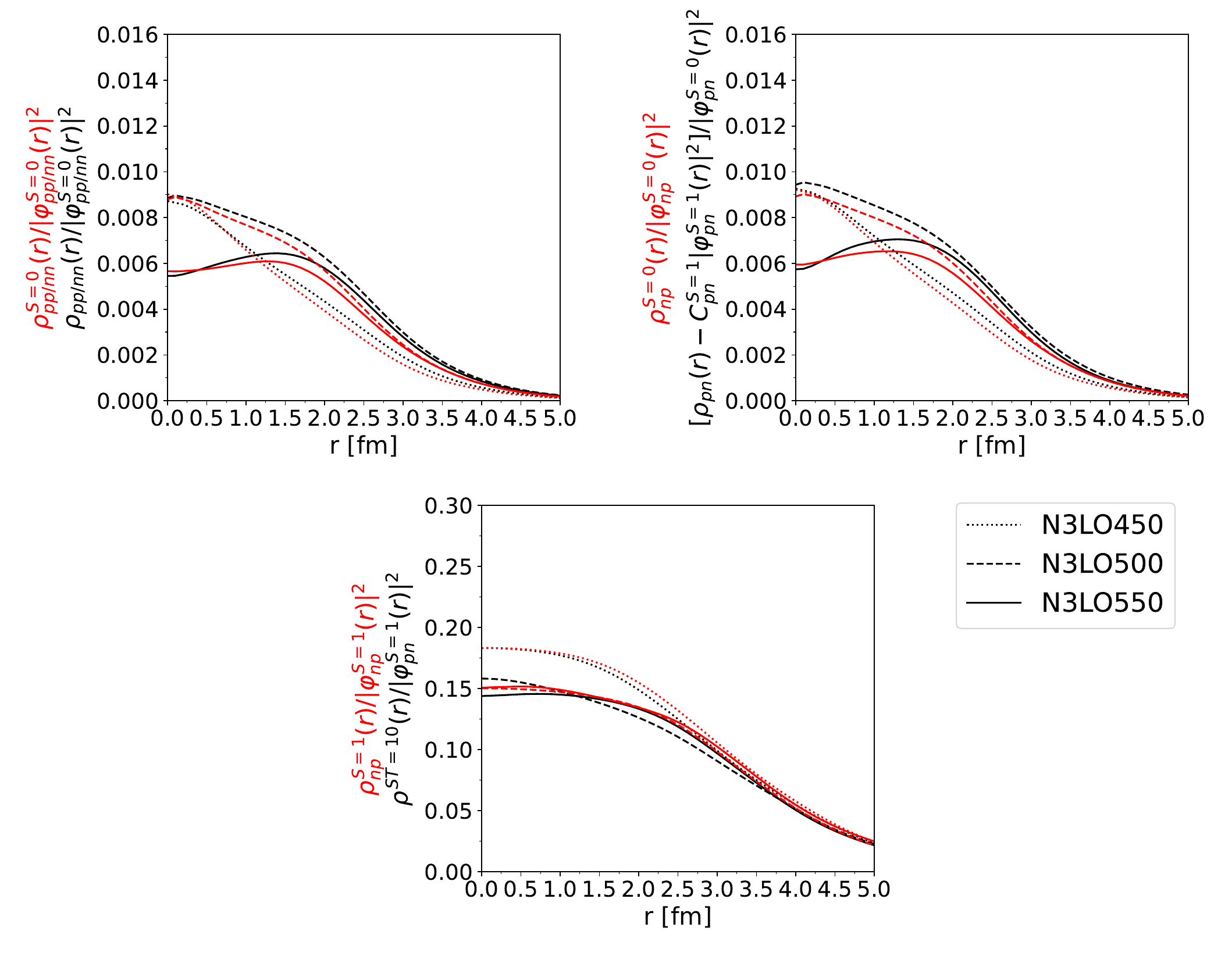}
\caption{Same as Fig.~\ref{fig:2bdf_ratio_nonloc_n3lo}, but for the $^4$He nucleus.}
\label{fig:2bdf_ratio_nonloc_n3lo_4he}
\end{figure*}
Regarding the 2BDFs ratios, the trends observed for $ ^3\text{He} $ is present also for $^4$He. We show in Figs.~\ref{fig:2bdf_ratio_loc_4he}
and~\ref{fig:2bdf_ratio_nonloc_n3lo_4he} the 2BDFs ratios to the universal functions obtained with local and non-local potentials (in the latter case, the N3LO at $\Lambda=450,500$ and 550 MeV, as representative case).
Small differences can be observed respect to the $^3$He case, as, for instance, the fact that the plateaus in the $ S = 0 $ channels are less pronounced, and the fact that for $\Lambda=450$ MeV (550 MeV) we observe an increasing (decreasing) trend for $r\to 0$. This trend, already seen for $^3$He (see Figs.~\ref{fig:2bdf_ratio_nonloc_n2lo} and~\ref{fig:2bdf_ratio_nonloc_n3lo}), is 
much more evident and pronounced for $ ^4\text{He} $ case (see Fig.~\ref{fig:2bdf_ratio_nonloc_n3lo_4he}).

In conclusion, we can say that the results for $ ^4\text{He} $ indicate that the $ L=0 $ approximation is less accurate compared to $ ^3\text{He} $, with more noticeable differences between the two calculations. As a consequence, the plateaus in the ratios extracted from the 2BMDs are better defined when the $ L=0 $ approximation is not applied. This holds true for both local and non-local potentials, further reinforcing the need to go beyond this approximation to achieve more precise and stable extractions of the contact coefficients.

We finally have performed a similar analysis of the TNI contribution also in the $A=4$ case. We do not report the relative figures, quite similar to the $^3$He case, except for Fig.~\ref{fig:2bmd_ratio_n3lo_noTNI_4he}, where we show the 2BDFs ratios for the non-local potentials at N3LO, with $\Lambda=450,500$ and 550 MeV, as a representative case. The conclusion of our analysis is that the TNI contributions remain relatively small, but slightly larger than in the $^3$He case.
\begin{figure*}[htbp]
\centering
\includegraphics[width=\linewidth]{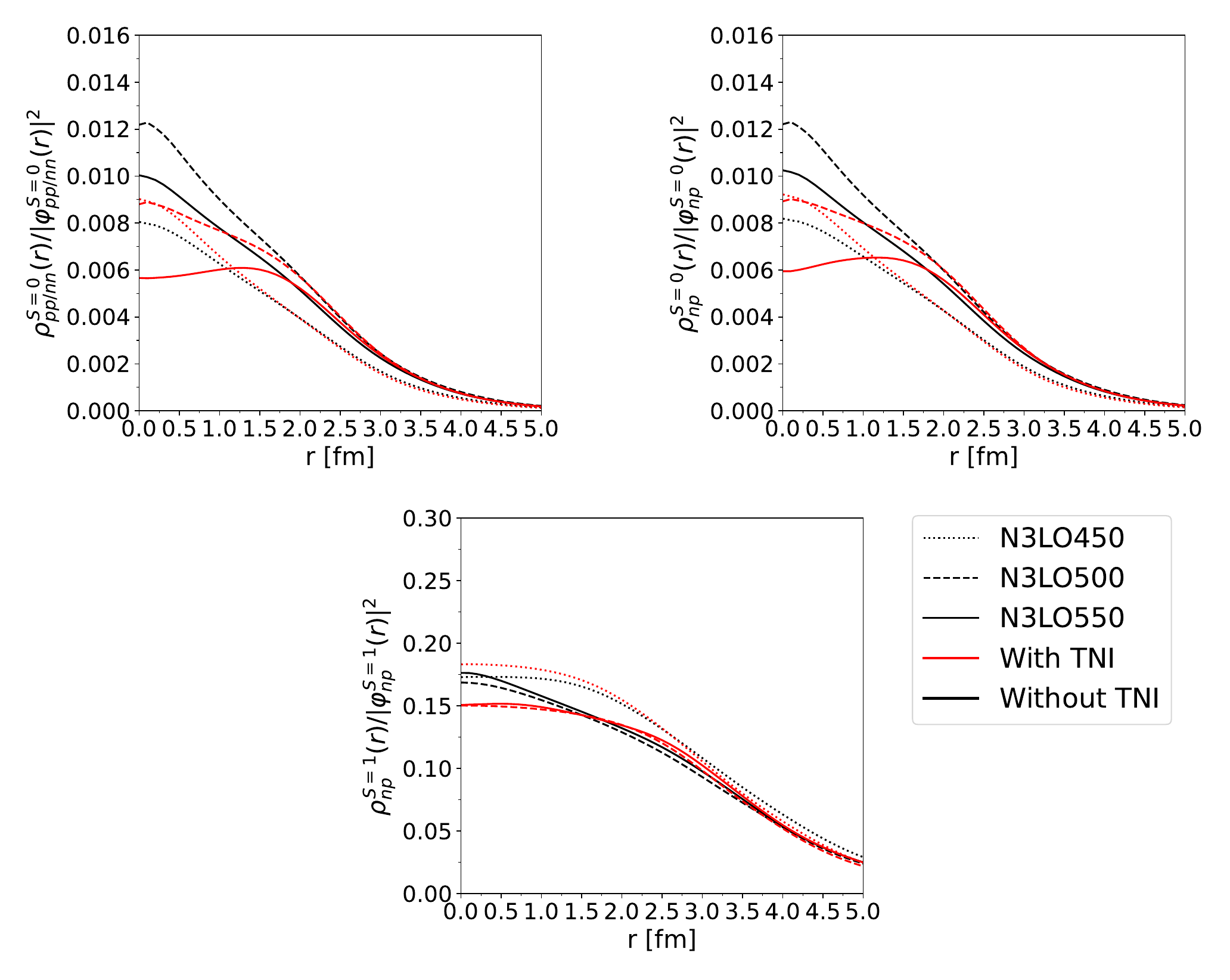}
\caption{Same as Fig.~\ref{fig:2bdf_ratio_n3lo_noTNI}, but for the $^4$He nucleus.}
\label{fig:2bmd_ratio_n3lo_noTNI_4he}
\end{figure*}

\subsection{$A=3,4$ contact coefficients}
\label{subsec:A34CC}

In this subsection, we present the values of the contact coefficients extracted as defined in Eqs.~\eqref{eq:cpp0_k}--\eqref{eq:cnp0_k} and Eqs.~\eqref{eq:cpp0_r}--\eqref{eq:cnp0_r}, with the procedure described in Sec.~\ref{subsec:NCC}, calculated using all the potentials listed in Sec.~\ref{subsec:2BDFMD}. As previously mentioned, in the GCF, the construction that connects short-distance and high-momentum physics leads to the expectation that the contact coefficients satisfy the relation $ \tilde{C}_A^{N_1 N_2, \alpha} = {C}_A^{N_1 N_2, \alpha} $. We will be able to verify this hypothesis with the largest variety of nuclear potentials.

\begin{figure*}
    \centering
    \includegraphics[width=1.0\textwidth]{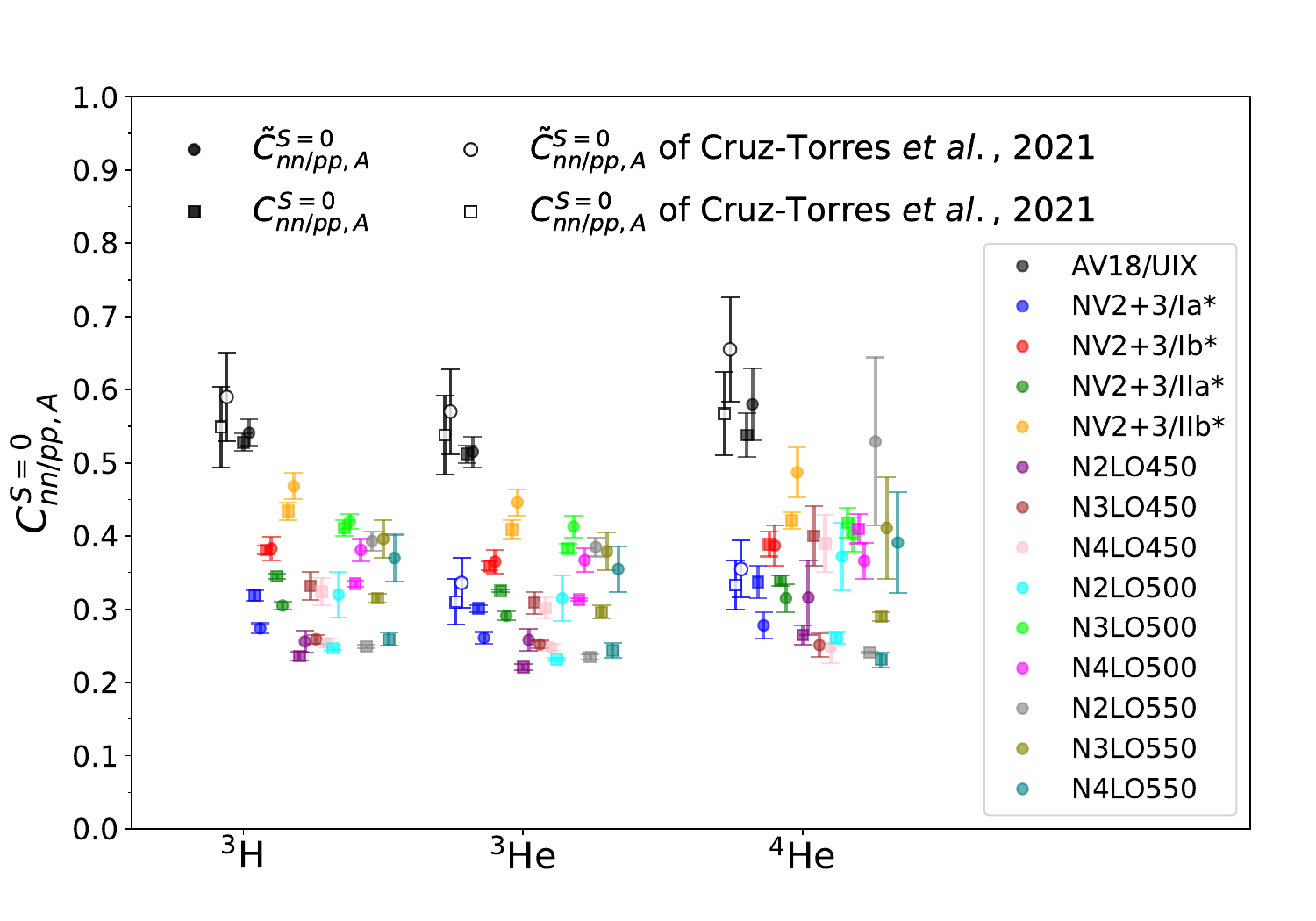}
    \caption{The contact coefficients 
    $\tilde{C}_{nn/pp, A}^{S=0}$ (circles) and $C_{nn/pp, A}^{S=0}$ (squares) for the $^{3}\text{H}$, $^{3}\text{He}$, and $^{4}\text{He}$ nuclei, calculated using the nuclear interaction models discussed in Sec.~\ref{subsec:2BDFMD}. Empty circles and empty squares represent the values of $\tilde{C}_{nn/pp, A}^{S=0}$ and $C_{nn/pp, A}^{S=0}$ obtained in Ref.~\cite{Cruz-Torres:2019fum} (labelled as Cruz-Torres {\it et al.}, 2021) for the AV18/UIX and NV2+3/Ia* potentials, the latter available only for $^{3}\text{He}$ and $^{4}\text{He}$.}
    \label{fig:cnn0}
\end{figure*}
\begin{figure*}[htbp]
\centering
\includegraphics[width=1\linewidth]{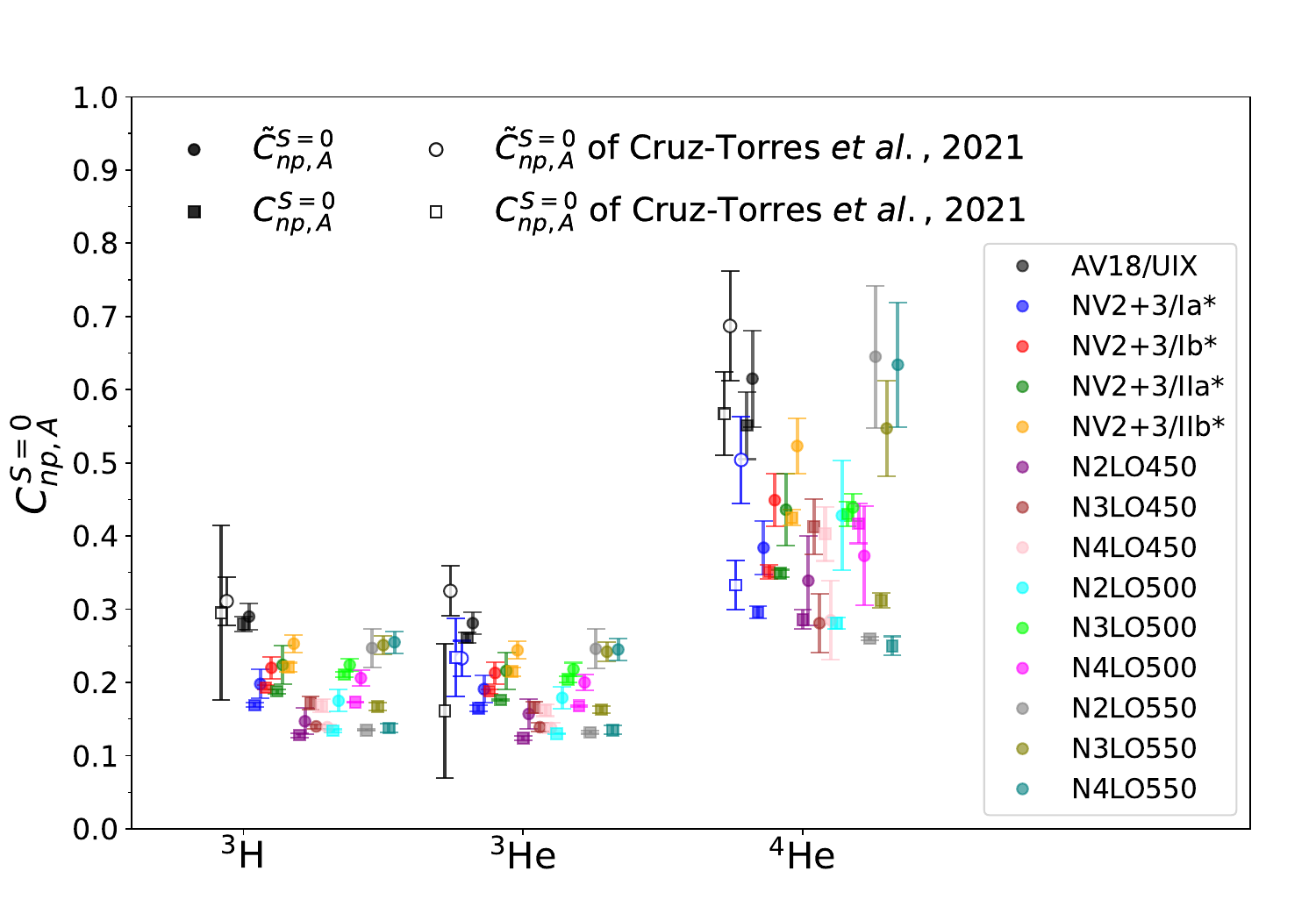}
\caption{Same as Fig.~\ref{fig:cnn0}, but for
$\tilde{C}_{np, A}^{S=0}$ and $C_{np, A}^{S=0}$.}
\label{fig:cnp0}
\end{figure*}
\begin{figure*}[htbp]
\centering
\includegraphics[width=1\linewidth]{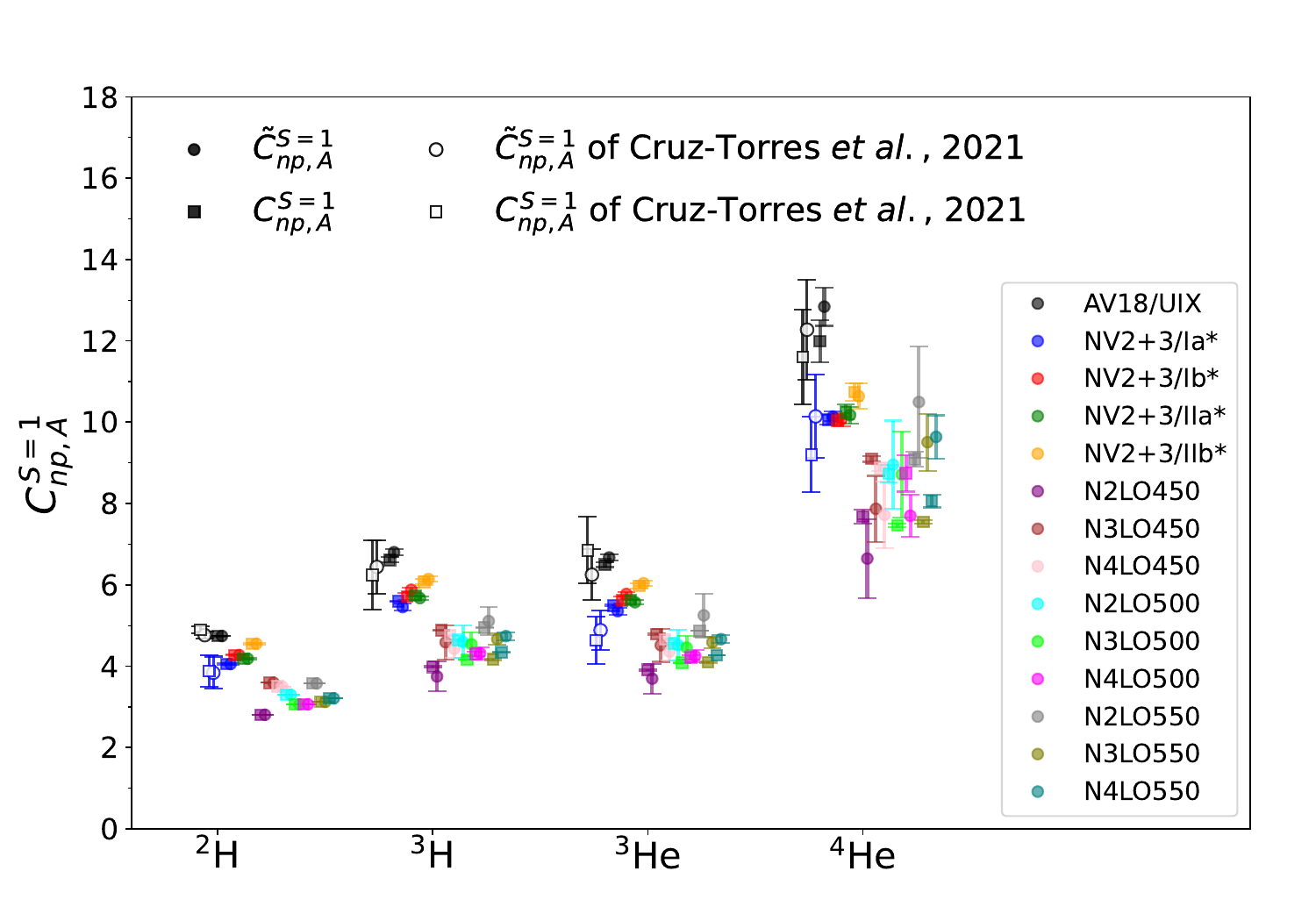}
\caption{Same as Fig.~\ref{fig:cnn0}, but for $\tilde{C}_{np, A}^{S=1}$ and $C_{np, A}^{S=1}$ for $^{2}\text{H}$, $^{3}\text{H}$, $^{3}\text{He}$, and $^{4}\text{He}$. }
\label{fig:cnp1}
\end{figure*} 
In Figs.~\ref{fig:cnn0}--\ref{fig:cnp1} we provide the results for the extracted contact coefficients calculated from the 2BMDs (circles) and the 2BDFs (squares). Note that coefficients calculated with the same potential are represented with the same color. 
In particular, in Fig.~\ref{fig:cnn0}, we present the values obtained for the $nn/pp$ contact coefficients of $ ^3\text{H} $, $ ^3\text{He} $, and $ ^4\text{He} $. By inspection of the figure, we can see that the results obtained with local potentials exhibit small errors, and lead to $\tilde{C}_{nn/pp, A}^{S=0}$ and $C_{nn/pp, A}^{S=0}$ in reasonable agreement across the three nuclei under investigation, confirming the expected behavior from the GCF. 
In contrast, the results obtained with non-local potentials exhibit larger errors, especially for the coefficients extracted from the 2BMDs,  a consequence of the fact that sometimes the plateau is not well defined. Differences between $\tilde{C}_{nn/pp, A}^{S=0}$ and $C_{nn/pp, A}^{S=0}$  are visible for many non-local potentials, especially those with cutoff $ \Lambda = 550 $ MeV. The errors are particularly pronounced for $ ^4\text{He} $, where the plateaus are less well-defined compared to the $ A=3 $ nuclei. This disagreement, with $\tilde{C}_{nn/pp, A}^{S=0}$ and $C_{nn/pp, A}^{S=0}$ differing in some cases by factors of 2 or 3, poses severe questions to the GCF.
In Fig.~\ref{fig:cnn0} we report also 
the contact coefficients obtained in Ref.~\cite{Cruz-Torres:2019fum} for the AV18/UIX and NV2+3/Ia* potentials, with the latter available only for $ ^3\text{He} $ and $ ^4\text{He} $. We observe excellent agreement with our results for the same potentials, but the results of Ref.~\cite{Cruz-Torres:2019fum} have much larger uncertainties. This is due to the fact that these results are calculated using the VMC method to solve for the $A=3,4$ bound states, and this method is less accurate than the HH method used here. Furthermore, the results from Ref.~\cite{Cruz-Torres:2019fum} were obtained with the $L=0$ assumption. Although the differences are very small for local potentials, they may still be visible to a minimal extent, especially for $ ^4\text{He} $.
The conclusion regarding the $S=0$ $nn/pp$ channel is that using local potentials, the GCF prediction that $\tilde{C}_{nn/pp, A}^{S=0} = C_{nn/pp, A}^{S=0}$ is reasonably satisfied across all three nuclei. For non-local potentials, however, this behavior holds only for a limited number of cases, with larger discrepancies observed for $ ^4\text{He} $. This represents a significant challenge for the GCF.

In Fig.~\ref{fig:cnp0}, we present the values obtained for the $np$ $S=0$ contact coefficients for the $ ^3\text{H} $, $ ^3\text{He} $, and $ ^4\text{He} $ nuclei, calculated using all the available potentials. By inspection of the figure, we can draw similar conclusions as in the $nn/pp$ $S=0$ case. In particular, results obtained with 
local potentials exhibit smaller errors than those obtained with the non-local ones. Furthermore, the GCF prediction that $\tilde{C}_{np, A}^{S=0} = C_{np, A}^{S=0}$ is fulfilled in this case. The results obtained with non-local potentials are more spread and in some cases $ \tilde{C}_{np}^{S=0} $ is significantly larger than $ C_{np}^{S=0} $ for the same potential, with the largest discrepancies observed for $ ^4\text{He} $. Specifically, for some non-local potentials, the values extracted from the 2BMDs are nearly twice those obtained from the 2BDFs. For all nuclei, the largest discrepancies occur for the $ \Lambda = 550 $ MeV cutoff, consistently with the trends observed in the $ nn/pp \, S=0 $ channel.
As in Fig.~\ref{fig:cnn0}, also here we report the contact coefficients obtained in Ref.~\cite{Cruz-Torres:2019fum} for the AV18/UIX and NV2+3/Ia* potentials. The agreement with our results is also in this case quite nice, although the results of Ref.~\cite{Cruz-Torres:2019fum} present in this channel a larger error, compared with the $nn/pp$ $S=0$ one. Therefore, conclusions similar to the $nn/pp$ $S=0$ channel can be drawn also for the $np$ $S=0$ one. 

In Fig.~\ref{fig:cnp1}, we present the values obtained for the $np$ $S=1$ contact coefficients for $ ^2\text{H} $, $ ^3\text{H} $, $ ^3\text{He} $, and $ ^4\text{He} $, using the same set of potentials.
By inspection of the figure, we can observe that for $ A=2 $, there are no errors associated with the extracted coefficients, since the ratio between the 2BDFs and the universal function is flat for any local or non-local potential. The ratio between the 2BMDs and the FT of the universal function, instead, is completely flat for local potentials, but not for non-local potentials at large values of $ k $, where cutoff effects emerge, as seen for $ A=3 $ and $ A=4 $. Thus, this region was excluded in our contact coefficients extraction. A second observation for deuteron is that for all potentials, the values of $ C_{np}^{S=1} $ and $ \tilde{C}_{np}^{S=1} $ are in perfect agreement.
For $ A=3 $ and $A=4$ nuclei, we again find that for both local and non-local potentials $\tilde{C}_{np, A}^{S=1} = C_{np, A}^{S=1}$, although  for non-local potentials the errors are larger.
For $ A=4 $, only in the case of the non-local potentials with cutoff $ \Lambda = 550 $ MeV, there are very small differences between 
 ${C}_{np}^{S=1} $ and  $ \tilde{C}_{np}^{S=1} $. Therefore, we can conclude that in the $ np$ $S=1 $ channel, which is the most significant one, we observe the GCF predicted behavior within our theoretical errors, for all nuclei and all potentials. Minor discrepancies are present for non-local potentials, especially for $ ^4\text{He} $. Note that also in Fig.~\ref{fig:cnp1} we report the results obtained in Ref.~\cite{Cruz-Torres:2019fum}, and we again observe that the errors are much larger than in our case. This is more evident in this channel because the extraction errors are very small, given that the plateaus are very well-defined. In any case, also in this case, we find a good agreement between these results and the ones presented here.

\subsection{Contact coefficients ratio} 
\label{subsec:A34CC-ratio}

We analyze in this subsection the ratio of contact coefficients, as described in Sec.~\ref{subsec:A34CC-ratio} and given in Eqs.~\eqref{ratio_r} and~\eqref{ratio_k}. We remark that, within the GCF framework, this ratio is expected to be the same, whether we use $C^S_{N_1 N_2,A}$ or $\tilde{C}^S_{N_1 N_2,A}$. In Ref.~\cite{Cruz-Torres:2019fum} this was shown for a large variety of nuclei, but for a limited set of local potentials. In our study, we are limited to $A=2,3,4$ nuclei, but we use a large variety of local and non-local interactions. 

\begin{figure*}
    \centering
    \includegraphics[width=\textwidth]{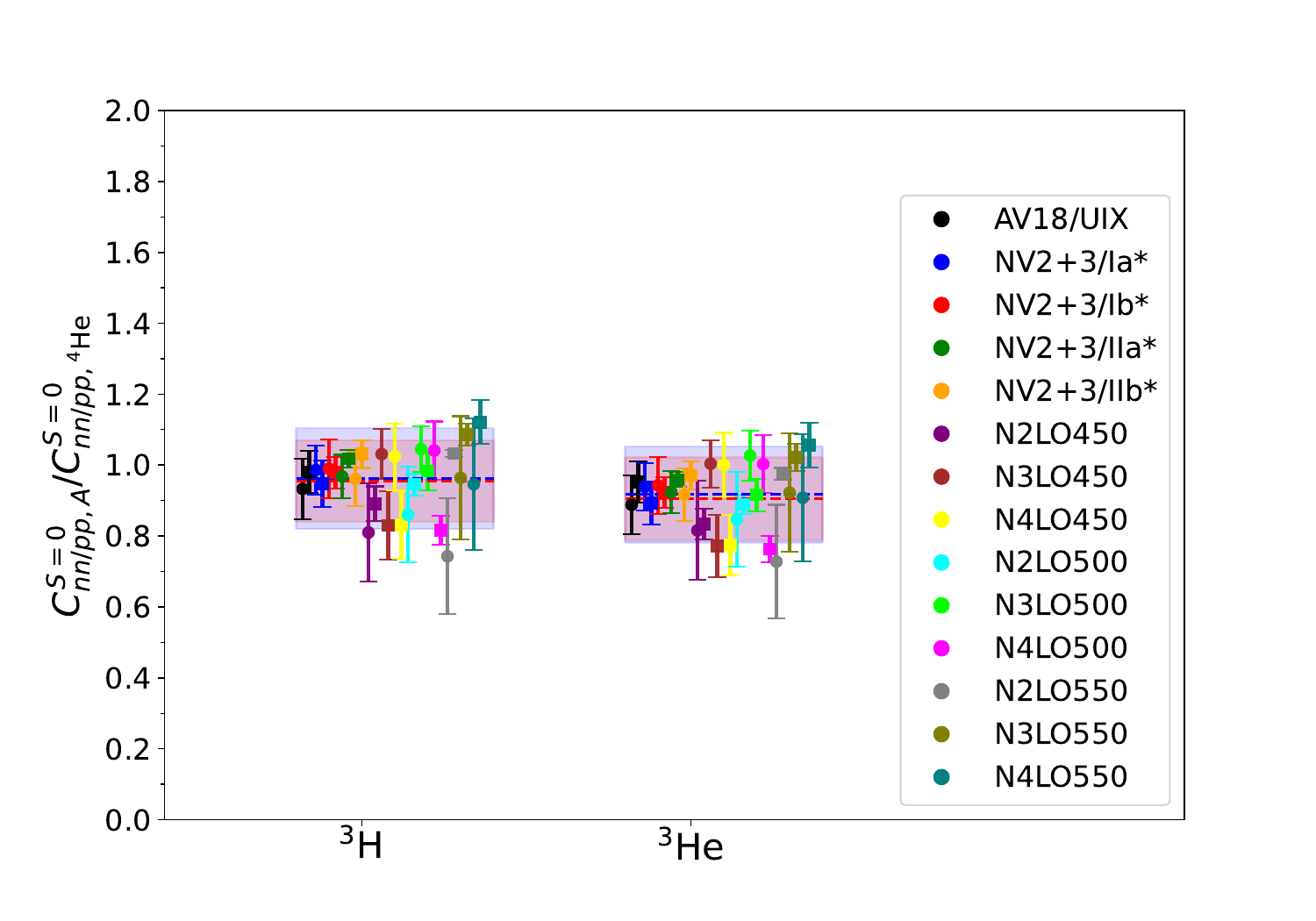}
    \caption{The contact coefficient ratios
    $\tilde{C}_{nn/pp, A=3}^{S=0}/ \tilde{C}_{nn/pp, ^4\text{He}}^{S=0} $ (circles) and $C_{nn/pp, A=3}^{S=0}/C_{nn/pp, ^4\text{He}}^{S=0}$ (squares) for the $^{3}\text{H}$ and $^{3}\text{He}$ nuclei, calculated using the nuclear interaction models discussed in Sec.~\ref{subsec:2BDFMD}. The ratios were fitted using a weighted average with four different sets of weights as discussed in Sec.~\ref{subsec:A34CC-ratio}. The blue and red dashed lines and bands represent the results obtained with Criterium 1 and 4 of Table~\ref{tab:ratios}, respectively.}
\label{fig:nn0ratio}
\end{figure*}

\begin{figure*}[htbp]
\centering
\includegraphics[width=1\linewidth]{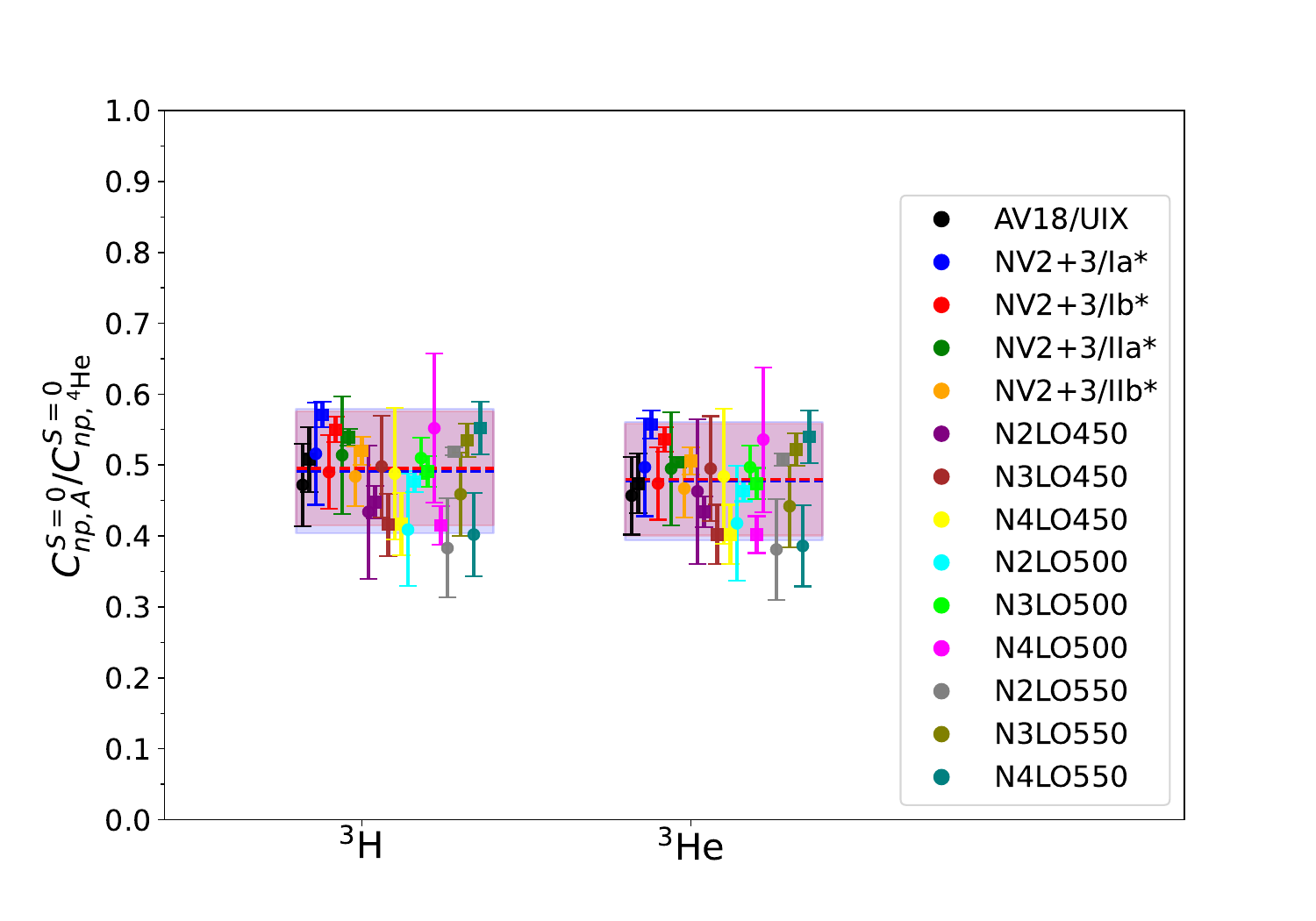}
\caption{Same as Fig.~\ref{fig:nn0ratio}, but for
 $\tilde{C}_{np, A=3}^{S=0}/ \tilde{C}_{np, ^4\text{He}}^{S=0} $ and $C_{np, A=3}^{S=0}/C_{np, ^4\text{He}}^{S=0}$.}
\label{fig:np0ratio}
\end{figure*}

\begin{figure*}
    \centering
\includegraphics[width=1\linewidth]{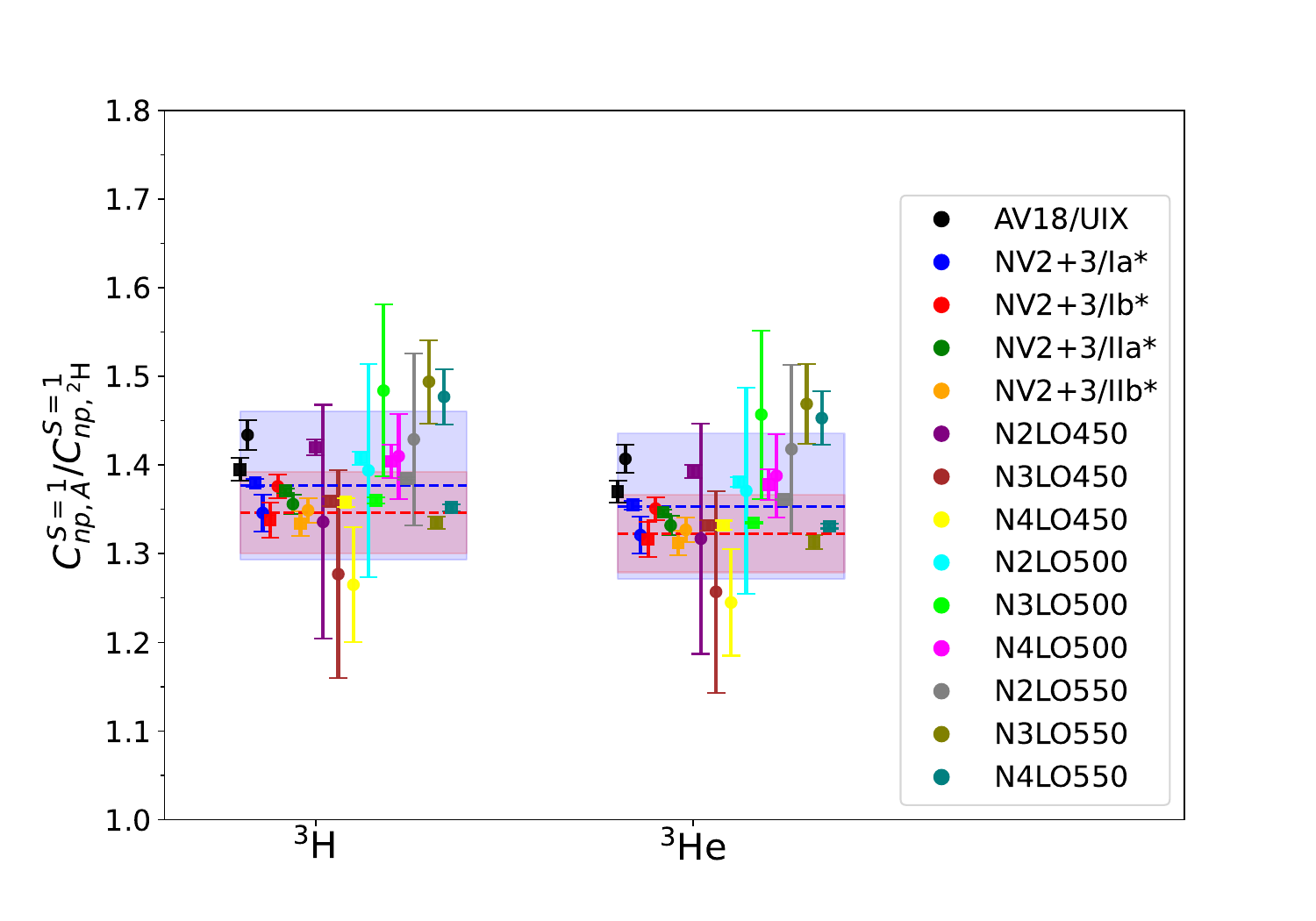}
\caption{Same as Fig.~\ref{fig:nn0ratio}, but for \ $\tilde{C}_{np, A=3}^{S=1}/ \tilde{C}_{np, ^2\text{H}}^{S=1} $ and $C_{np, A=3}^{S=1}/C_{np, ^2\text{H}}^{S=1}$.}
\label{fig:np1ratio-3}
\end{figure*}

\begin{figure*}
    \centering
\includegraphics[width=1\linewidth]{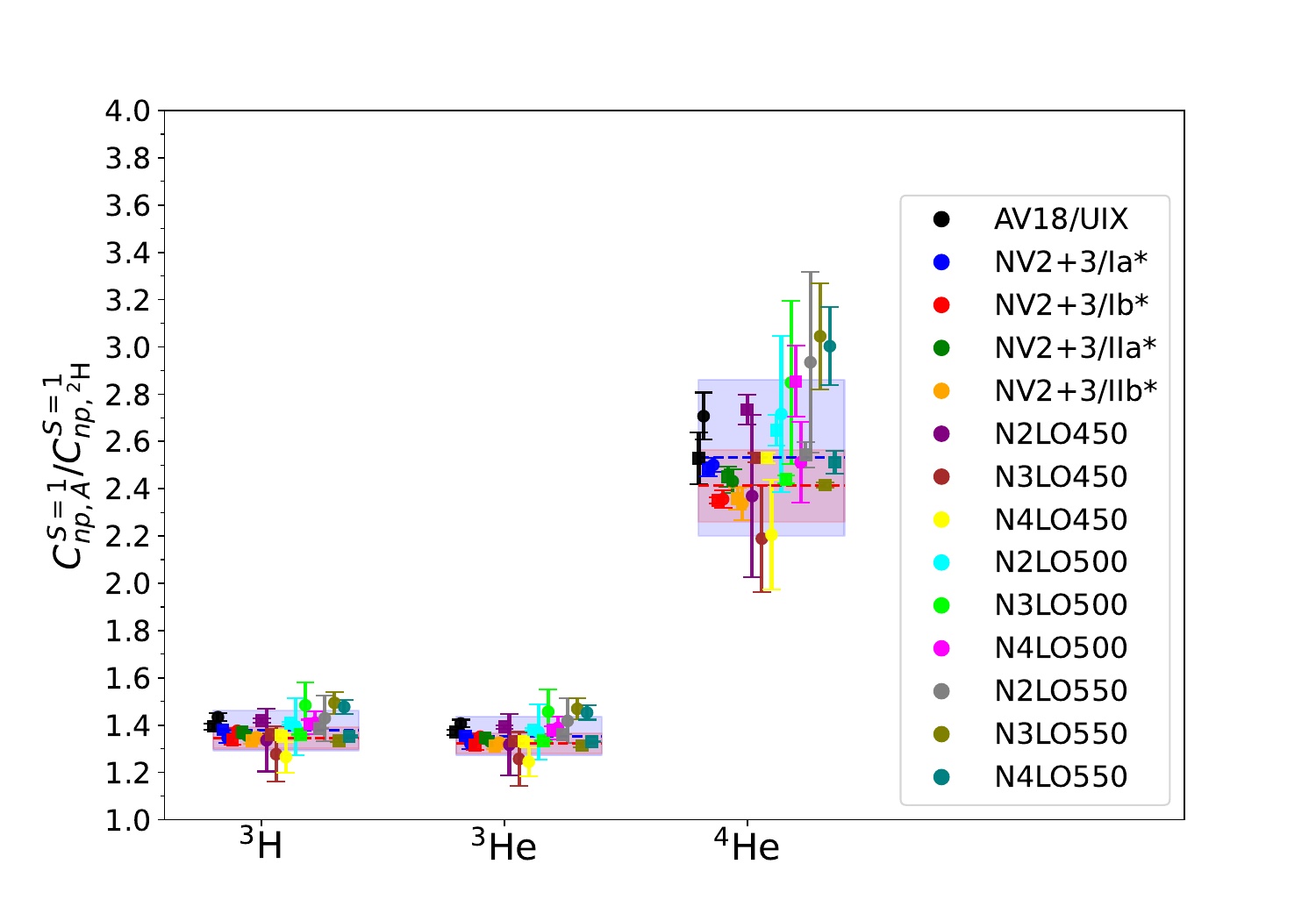}
\caption{Same as Fig.~\ref{fig:np1ratio-3}, but for the $^{3}\text{H}$, $^{3}\text{He}$, and $^{4}\text{He}$ nuclei.}
\label{fig:np1ratio-34}
\end{figure*}

In Figs.~\ref{fig:nn0ratio}--\ref{fig:np1ratio-3}, we report the ratios between the contact coefficients for $A=3$ nuclei in the $nn/pp$ $S=0$, $np$ $S=0$ and $np$ $S=1$ channels, respectively. In Fig.~\ref{fig:np1ratio-34} we report the ratios in the $np$ $S=1$ channel for $ ^3\text{H} $, $ ^3\text{He} $, and $ ^4\text{He} $. By inspection of the figures, we notice first of all that the uncertainties associated with local potentials are very small, whereas non-local potentials exhibit significantly larger errors for the ratios $\tilde{C}^S_{N_1 N_2,A}/\tilde{C}^S_{N_1 N_2,A_0}$, as expected from Figs.~\ref{fig:cnn0}--\ref{fig:cnp1}.
Additionally, we observe that the ratios between the coefficients extracted from the 2BMDs and those extracted from the 2BDFs are in good agreement. This holds not only for local potentials, but also for non-local potentials, for which, in contrast to the local potentials results, the individual coefficients deviated from the GCF prediction -- especially for the $ S = 0 $ channels.
Only in the $ np $ $ S = 0 $ channel, small differences are observed, especially for the non-local potentials with cutoff $ \Lambda = 550 $ MeV.
Note that the results obtained with the  $ \Lambda = 550 $ MeV potentials present minor differences also in the other two channels.

Figs.~\ref{fig:nn0ratio}--\ref{fig:np1ratio-34} allow to address 
the model-independence of the contact coefficients ratios in the various spin-isospin channels. In particular, we can see that for local potentials, model-independence is easily obtained in all channels, as the ratios are very close to each other. The ratio for the phenomenological AV18/UIX potential is slightly larger compared to those calculated with the local chiral potentials in all channels, but the differences are very small.
For non-local potentials, however, a significantly larger dispersion is observed in all channels. This behavior is particularly pronounced in the channels with $ S = 0 $ compared to the dominant  $ np$  $S = 1 $ channel. Despite relatively large uncertainties, in the latter channel, only some potentials show a more significant dispersion in the ratios calculated from the contact coefficients (see Figs.~\ref{fig:np1ratio-3} and \ref{fig:np1ratio-34}).
Nevertheless, model-independence can still be advocated when considering both local and non-local potentials.

In order to be able to provide our best estimate for the vaious ratios with a robust assigned theoretical uncertainty, we defined for each nucleus and each spin-isospin channel, the avarage and variance as~\cite{Gnech:2023mvb}
\begin{eqnarray} 
\langle C \rangle &=& \sum_i C_i P_i\ , \label{eq:c-avarage}\\
\delta^2 \langle C \rangle &=& \delta C_i P_i + \delta^2 C_{\text{syst}}\ ,\label{eq:delta2c} \\
\delta^2 C_{\text{syst}} &=& \left[\sum_i C_i^2 P_i\right] - \left[\sum_i C_i P_i\right]^2 \ ,\label{eq:deta2c-syst} 
\end{eqnarray}
where the index $i=1 \ldots N$ runs on the various potentials (see below), $C _i$ are the contact coefficient ratios, $\delta C_i$ the associated errors, and $P_i$ are weights assigned to each case. The choice of these weights is clearly arbitrary. However, in order to show the stability of our results, we assigned $P_i$ according to 4 different criteria, listed below.

\begin{itemize}
    \item{Criterium 1}:  
    We separate the results in two groups, depending on whether they are obtained with the local or non-local potentials, since the latter exhibit a more dispersive behavior, assigning an equal weight of $\frac{1}{2}$ to both groups. Since we have used 5 local and 9 non-local potentials, the weights $P_i$ are chosen as 
\begin{equation*}
P_i = 
\begin{cases} 
\frac{1}{10}, & \text{for local potentials,}  \\
\frac{1}{18}, & \text{for non-local potentials.} 
\end{cases}
\end{equation*}
In this case, $N=14$.

    \item{Criterium 2}: 
    The second choice for $P_i$ starts from the observation that of the 14 adopted potentials, one is purely phenomenological (AV18/UIX), primarily used to benchmark our results against those of Ref~\cite{Cruz-Torres:2019fum}, and 13 are chiral potentials, divided into local and non-local ones. If we decide to provide our best estimate for the contact coefficients ratios based only on chiral potentials, we need to exclude the AV18/UIX results. Then, the weights $P_i$ are chosen as
    \begin{equation*}
    P_i = 
    \begin{cases} 
    \frac{1}{8}, & \text{for local chiral potentials,}  \\
    \frac{1}{18}, & \text{for non-local chiral potentials.} 
    \end{cases}
    \end{equation*}
In this case, $N=13$.

    \item{Criterium 3}:     
    A further choice, which differs from the previous ones, is to treat all the potentials equally, and therefore we can assign the same weight to all the various cases, i.e.\ 
    \begin{equation*}
    P_i = \frac{1}{14}, \quad \text{for each potential.}
    \end{equation*}
Obviously, in the case, $N=14$ again.

    \item{Criterium 4}: 
    Finally, we consider again only chiral potentials and separate them into local and non-local ones. However, since the local Norfolk potentials are calculated at fixed chiral order (N3LO), we restrict our selection for the non-local chiral potentials to the same order. Consequently, we are left with the three non-local potentials with cutoff $\Lambda=450,500$ and $550$ MeV. The weights $P_i$ then are chosen as
    \begin{equation*}
      P_i = 
    \begin{cases} 
    \frac{1}{8}, & \text{for local Norfolk potentials,}  \\
    \frac{1}{6}, & \text{for non-local N3LO potentials.} 
    \end{cases}   
    \end{equation*}
In this case $N=7$.
\end{itemize}

The contact coefficients ratios in the various channels, applying the various criteria are listed in Table~\ref{tab:ratios}. 
\begin{table*}[htbp]
\begin{tabular}{|c|c|c|c|c|}
\hline
$nn/pp$ $S=0$ & Criterium 1 & Criterium 2 & Criterium 3 & Criterium 4 \\
\hline
$^3$H  & 0.978 $\pm$ 0.147 & 0.981 $\pm$ 0.147 & 0.978 $\pm$ 0.162 & 0.968 $\pm$ 0.119 \\
$^3$He & 0.932 $\pm$ 0.140 & 0.934 $\pm$ 0.139 & 0.933 $\pm$ 0.153 &  0.918 $\pm$ 0.123 \\
\hline\hline
$np$ $S=0$ & Criterium 1 & Criterium 2 & Criterium 3 & Criterium 4 \\
\hline
$^3$H  & 0.496 $\pm$ 0.083 & 0.499 $\pm$ 0.084 & 0.490 $\pm$ 0.086 & 0.501 $\pm$ 0.072 \\
$^3$He & 0.481 $\pm$ 0.079 & 0.485   $\pm$ 0.080 & 0.477 $\pm$ 0.082 & 0.485 $\pm$ 0.071 \\
\hline\hline
$np$ $S=1$ & Criterium 1 & Criterium 2 & Criterium 3 & Criterium 4 \\
\hline
$^3$H  & 1.377 $\pm$ 0.084 & 1.371 $\pm$ 0.082 & 1.380 $\pm$ 0.092 & 1.347 $\pm$ 0.046 \\
$^3$He & 1.353 $\pm$ 0.082 & 1.348 $\pm$ 0.081 & 1.356 $\pm$ 0.090 & 1.323 $\pm$ 0.044 \\
$^4$He & 2.520 $\pm$ 0.332 & 2.499 $\pm$ 0.329 & 2.540 $\pm$ 0.360 & 2.400 $\pm$ 0.143 \\
\hline
\end{tabular}
\caption{\label{tab:ratios} Contact coefficients ratios in the various channels, for $A=3,4$ nuclei, applying the various criteria defined in the text.}
\end{table*}
By inspection of the table, we can conclude that the criteria 1--3 lead to very similar results, slightly different from those obtained with criterium 4. These differences are however very small, and well within the associated errors. In fact, thanks to our fitting procedure, we can conclude that 
the contact coefficients ratios are all compatible, with the uncertainty assigned to the mean value according with Eqs.\eqref{eq:c-avarage}--\eqref{eq:deta2c-syst} for most of the potentials, both local and non-local. Even in cases where discrepancies are observed, these are very small. Consequently, the final values obtained for the contact coefficients ratios, along with their uncertainties, across the different channels and nuclei, quantify model-dependence, i.e.\ we can assume model-independence is satisfied within our estimated theoretical uncertainty. 

\section{Conclusions and outlook}
\label{sec:conc-out}

In this work we have investigated SRCs in light nuclei, focusing on the behavior of nucleons at short distances or high momenta through the GCF. Specifically, we have applied the GCF framework to extract the nuclear contact coefficients from 2BDFs and 2BMDs, using a large variety of interaction models, including both local and non-local potentials, and focusing on $A = 2, 3,$ and $4$ nuclei. We studied the contact coefficients across different spin channels, i.e. $nn/pp$  $S=0$, $np$ $S=0$, and $np$  $S=1$.

Our first conclusion regards the $ L = 0 $ approximation, used in previous studies~\cite{Cruz-Torres:2019fum,WEISS2018211} to extract the contact coefficients. In particular, we have observed that while the differences between the calculations with and without the $ L = 0 $ approximation are evident, they are generally small in the plateau region of the 2BDFs and 2BMDs ratios with the universal functions and their FT. However, in some cases, differences between the results obtained with or without the $L=0$ approximation are more pronounced, as for the $^4$He nucleus. This highlights the importance of going beyond the $ L = 0 $ approximation, as it typically leads to a more accurate identification of the plateau regions, for both local and non-local potentials. 

Our second conclusion regards the role of TNI. In fact, TNI contributions are usually small, especially in the $np$ $S=1$ channel. Only in the $S=0$ channels, the TNI contributions in the plateau regions of the 2BDFs ratios, especially for non-local potentials, are quite pronounced, particularly for $ ^4\text{He} $. 

A third conclusion regards the GCF prediction that 
$\tilde{C}_{N_1 N_2, \, A}^S = C_{N_1 N_2, \, A}^S$ for all spin channels, for a given potential and nucleus~\cite{Cruz-Torres:2019fum,WEISS2018211}.
While this is reasonably verified in the 
dominant $np$ $S=1$ channel, for all nuclei and all potentials, in the $S=0$ channels, for non-local potentials, particularly with cutoff $\Lambda = 550$ MeV, we observed significant differences between $\tilde{C}_{N_1 N_2,A}^{S=0}$ and $C_{N_1 N_2,A}^{S=0}$, especially for $^4\text{He}$. In some cases, for the $np$ $S=0$ channel, the value of $\tilde{C}_{np,A}^{S=0}$ is nearly twice as large as ${C}_{np,A}^{S=0}$. Additionally, the coefficients extracted from the 2BMDs for non-local potentials present larger errors, reflecting the poorly defined plateaus. As a result, while the GCF predictions are somehow confirmed for local potentials in the $S=0$ channels, they are very poorly satisfied for non-local potentials.This poses significant questions on the validity of the GCF at least in the $S=0$ channels.

Finally, we have explored the model-independence of the contact coefficient ratios across different spin-isospin channels, calculating a weighted average for each nucleus and channel. Four weighting schemes have been explored, to account for the different nature of the adopted interaction models, finding an overall nice agreement among the various schemes and among the various contact coefficient ratios. The values for these ratios are summarized in Table~\ref{tab:ratios}.

The results obtained in this work have important implications. 
First of all, they pose significant questions to the GCF, especially
on the non-dominant $S=0$ channels, where large differences have been found 
between $\tilde{C}_{N_1 N_2,A}^{S}$ and ${C}_{N_1 N_2,A}^{S}$, especially for
non-local potentials.
One possible cause of these discrepancies may stem from the specific form of the adopted NN potential, and, less likely, from a mismatch in the cutoff functions of the NN potential and the TNI. Further investigation of this aspect is currently underway. On the other hand, the large tensions between our results and the GCF predictions relax in the case of the contact coefficient ratios. Finally, we believe that the results reported in Table~\ref{tab:ratios} allow to estimate more reliably the theoretical uncertainties to be associated to GCF studies of contact coefficients ratios in heavier systems, of the order of at least 10-20\%. To this regard, we plan to extend our study to the $A = 6$ nuclei, still using the HH method~\cite{Gnech:2023mvb}, so that both local and non-local interactions will be studied.
    
\section*{Acknowledgements}
The Authors are very grateful to R.\ Weiss and R.\ Cruz-Torres for very useful discussions on the GCF and on the details of the calculation of Ref.~\cite{Cruz-Torres:2019fum}. The computing resources of the INFN-Pisa branch are also gratefully acknowledged.

\bibliography{bib}

\end{document}